%% file: dbgap.tex
\documentclass[preprint]{aastex}
\usepackage{lscape}
\usepackage{epsf}

\newcounter{affil}

\newcommand{\tableskip}{\\[-6pt]}

\newcommand{\doubleline}{\tableskip\hline\hline\tableskip}

\newcommand{\msun}{\ensuremath{{\rm\; M}_\odot}}

\newcommand{\mystar}[1]{{\bf #1:}}

\newcommand{\logg}{\ensuremath{\log g}}

\begin{document}
\title{Hot DB White Dwarfs from the Sloan Digital Sky Survey%
\footnote{Some observations reported here were obtained at the MMT Observatory, a joint facility of the University of Arizona and the Smithsonian Institution.}}
\author{Daniel J.~Eisenstein\altaffilmark{1,8}, 
James Liebert\altaffilmark{1}, 
Detlev Koester\altaffilmark{2}, 
S.J.\ Kleinmann\altaffilmark{3,4},
Atsuko Nitta\altaffilmark{3,4},
Paul S.~Smith\altaffilmark{1}, 
J.C.\ Barentine\altaffilmark{3},
Howard J.\ Brewington\altaffilmark{3},
J.\ Brinkmann\altaffilmark{3},
Michael Harvanek\altaffilmark{3},
Jurek Krzesi\'nski\altaffilmark{3,5},
Eric H.\ Neilsen, Jr.\altaffilmark{6},
Dan Long\altaffilmark{3},
Donald P.\ Schneider\altaffilmark{7}, 
Stephanie A.\ Snedden\altaffilmark{3}
}

\begin{abstract}
We present $ugriz$ photometry and optical spectroscopy for 28 DB and DO
white dwarfs with temperatures between 28,000K and 45,000K.  About 10
of these are particularly well-observed; the remainder are candidates.
These are the hottest DB stars yet found, and they populate the ``DB gap''
between the hotter DO stars and the familiar DB stars cooler than 30,000K.
Nevertheless, after carefully matching the survey volumes, we find
that the ratio of DA stars to DB/DO stars is a factor of 2.5 larger at
30,000~K than at 20,000~K, suggesting that the ``DB gap'' is indeed
deficient and that some kind of atmospheric transformation takes place
in roughly 10\% of DA stars as they cool from 30,000~K to 20,000~K.
\end{abstract}


\altaffiltext{1}{Steward Observatory, University of Arizona,
                933 N. Cherry Ave., Tucson, AZ 85121}
\altaffiltext{2}{Institut f\"ur Theoretische Physik und Astrophysik,
	University of Kiel, D-24098, Kiel, Germany}
\altaffiltext{3}{Apache Point Observatory,
                P.O.\ Box 59, Sunspot, NM 88349}
\altaffiltext{4}{Subaru Telescope, 650 N. A'Ohoku Place, Hilo HI, 96720}
\altaffiltext{5}{Mt. Suhora Observatory, Cracow Pedagogical University, 
	ul.\ Podchorazych 2, 30-084 Cracow, Poland}
\altaffiltext{6}{Fermilab National Accelerator Laboratory,
     P.O. Box 500, Batavia, IL 60510}
\altaffiltext{7}{Department of Astronomy and Astrophysics,
                Pennsylvania State University, University Park, PA 16802}
\altaffiltext{8}{Alfred P.~Sloan Fellow}

\section{Introduction}
\label{sec:intro}

Existing surveys of white dwarfs have shown a peculiar absence of
stars with helium atmospheres with temperatures between roughly 30,000K
and 45,000K.  Currently, the hottest published DB temperature is for
the star PG~0112+104 near 30,000~K \citep{lie86}.
The most accurate determination is probably
$T_{\rm eff} = 30,783\pm269$~K, $\logg = 7.78\pm0.02$ \citep{bea95} using fits to He~I lines.  At this
temperature and below, only He~I lines appear in the spectrum.  Hydrogen
often appears as a trace constituent, and these spectra are classified
DBA.  Beauchamp analyzed 51 DB and DBA stars, none hotter than PG~0112+104.  
\citet{The91} analyzed the far ultraviolet
spectrophotometry of 11 of the hottest known DB stars but generally
showed these to have lower $T_{\rm eff}$ than previously-published values.
More recently
\citet{koe01} analyzed 19 more DB stars, but the hottest $T_{\rm eff}$
determination was 25,400~K.

The He~II 4686\AA\ line should start appearing at about 40,000~K
(38,000~K for a high signal-to-ratio spectrum), so spectra at this
temperature and above would be classified DO.  However, the coolest
published temperature for a DO star before any publications from the
Sloan Digital Sky Survey (SDSS; York et al. 2000) is the
rather crude estimate of $47,500\pm2,500$~K for PG~1133+489
\citep{wes85}.  Unfortunately, this star was not included in the more
precise analysis of DO stars by \citet{dre96}.  These authors included
DO stars discovered in later surveys such as the Hamburg Schmidt survey
\citep{heb96}, but did not discover any cooler objects, thus supporting
the existence of a ``DB gap''.  One or two stars classified DO from the
SDSS may be similar in temperature to PG~1133+489 \citep{Krz04}, but
none of the stars discussed in that paper are considerably cooler.

One suggested cause of this ``DB gap'' has been that as the DO stars
cool, small residual amounts of hydrogen diffuse to their surfaces, so
that 50,000~K helium atmosphere stars become 40,000~K hydrogen DA white
dwarfs \citep{fon87}.  \citet{fon87} also proposed that the
reappearance of helium atmosphere stars below 30,000~K is due to
convective mixing of the H-rich atmosphere into the much more massive
helium envelope when the outer convection boundary reaches high enough
to penetrate the hydrogen.  An unpleasant aspect of this theory is that
the hydrogen layer mass evidently has to be quite thin, of the order of
$10^{-15}$ of the stellar mass, such 
as was found for PG~1305--017 \citep{ber94}.  
If the white dwarf progenitor's
asymptotic giant branch phase is not terminated with a special episode
of mass loss, the remaining hydrogen envelope is predicted by models to
be of the order $10^{-4}\msun$.  Asteroseismological studies of
several pulsating DAV white dwarfs appear to confirm that these objects
have hydrogen layer masses at least within a few orders of magnitude of
this high value \citep{cle95,bra98}.  Since the majority of white dwarfs
within the DB temperature range remain DA, these apparently conflicting
results may perhaps be consistent if a small fraction of
the DAs retain only ultrathin outer hydrogen layers that can be convectively
mixed below 30,000~K.

The problem of whether a DB gap exists is complicated by the known
existence of several peculiar DAB, DBA, or DAO stars believed to lie in
the 30--45,000~K range that (1) show evidence of spectrum variability
and/or (2) do not fit simple atmospheric models.  Simple atmospheric
models may be either homogeneous (completely mixed) in H and He throughout
the atmosphere or stratified with the hydrogen all in a very thin, upper
layer.  An example near 45,000~K is PG1210+533, which shows variable H,
He I, and He II line strengths, probably modulated with an (unknown)
rotational period like the magnetic star Feige 7 \citep{lie77,ach92}.
PG~1210+533 is not known to be magnetic.  An example near 30,000K
is GD~323, which shows a peculiar DAB spectrum that has not been fit
successfully with either a homogeneous or stratified model atmosphere
\citep{lie84,koe94}.  This star has a variable spectrum \citep{per05}
but is again not known to be magnetic.  In contrast to these are two 
stars in the middle of the gap which are
well fit with homogeneous atmospheres models. HS~0209+0832 is fitted
at 36000~K with a 2\% helium abundance \citep{Jor93} and
PG~1603+432 at 35,000K with 1\% helium \citep{Ven04}.
The peculiarity of some of these stars
might be related to the edges of the DB gap, but in any case their 
uncertain relationship to the DB sequence has led workers to exclude
them, perhaps too quickly, from the statistics of the DB gap.
The possible roles of stellar wind mass loss, accretion and convective
dredge-up in the spectral evolution of DAB stars is discussed in
\citet{Mac91}.

It must be emphasized, however, that the total number of well-analyzed DB
stars prior to the SDSS was rather small, only slightly more than the 70
listed in the first paragraph.  Thus, the validity and significance of
the alleged ``DB gap'' has not been clear.  However, SDSS has included
many such stars in its spectroscopic program, resulting in a substantial
increase in the census of DBs (as well as many more DAs).  Here we
report the discovery of 6 DB white dwarfs with high signal-to-noise
ratio photometry and spectroscopy that place their temperatures in the
previously empty range of 30,000~K to 45,000~K.  We also discuss 4 other
well-observed stars at the edge of the range, and 18 less well-observed
candidates also likely to be in the DB gap.

\section{The SDSS}
\label{sec:sdss}

The SDSS \citep{Yor00,Sto02,Aba03,Aba04,Aba05,Ade06} is imaging 10,000 square
degrees of high Galactic latitude sky in five passbands, $u$, $g$, $r$,
$i$, and $z$ \citep{Fuk96,Gun98,Gun05}.  The images are processed
\citep{Lup01,Sto02,Pie02} and calibrated \citep{Hog01,Smi02,Ive04,Tuc05} to produce
5-band catalogs, from which galaxies, quasars, and stars are selected
for follow-up spectroscopy.  Spectra covering 3800\AA\ to 9200\AA\ with
resolution of 1800 are obtained with twin fiber-fed
double-spectrographs.  High priority targets, namely the primary
galaxies and quasars samples \citep{Eis01,Str02,Ric02} as well as a
small number of stars, are assigned to plug plates with a tiling
algorithm that ensures nearly complete samples, save for the effects of
the $55''$ fiber collision distance \citep{Bla01t}.  Lower priority
targets take left-over fibers, generally where the large-scale structure
of galaxies produce voids; these samples are incomplete and spatially
inhomogeneous, but likely in a manner that can be modeled.

Of particular importance to the study of DB white dwarfs is the
``Hot Standard'' target class, which selects all isolated stars with
clean photometry flags with very blue colors, $u-g<0$ and $g-r<0$,
down to a flux limit of $g<19$.  Both the colors and flux limit are
applied to dereddened magnitudes \citep{SFD}, as is likely appropriate because
stars this hot and faint are generally several hundred parsecs out of the disk.
Because of the lack of a Balmer decrement, DBs hotter than
about 15,000K are blue enough to make this target class \citep{Har03,Kle04}.  
The Hot Standard class is a tiled target class, meaning that these objects
are guaranteed a fiber to the limits of fiber collisions.  Note that
the requirement that the stars be isolated, i.e. not superposed on the
sky with another object, and free from certain photometry flags means that even the Hot Standard
class is only 50\% complete \citep{Eis05}.  
Fainter and redder WDs can still receive fibers by several other target
classes; see \citet{Kle04} for discussion of this issue.  Isolated stars at $g<19.5$
and $u-g<0.7$ and $g-r<-0.1$ that are not Hot Standards have a spectroscopic
completeness that is about 66\% of that of the Hot Standards \citep{Eis05}.
All SDSS targeting classes are also subject to a bright flux limit; these
vary slightly, but generally require $g>15$, not corrected for extinction,
in a $3''$ diameter fiber magnitude,
which corresponds to about $g>15.3$ in the usual PSF magnitudes.  

The SDSS photometric zeropoints are close to the AB convention
of 3631 Jy for $m=0$; however, they are not perfect.  We have 
adopted corrections of 
$u_{\rm AB} = u_{\rm SDSS} -0.04$, 
$i_{\rm AB} = i_{\rm SDSS} +0.015$, 
$z_{\rm AB} = z_{\rm SDSS} +0.03$, with $g$ and $r$ unchanged.
These match the value in \citet{Eis05} but differ somewhat from the
estimates in \citet{Aba04}.  The AB corrections have been estimated
from comparisons with the HST calibrations \citep{Boh01} and from
the DA color locus.
The $z$ band shift is uncertain but has negligible consequence for this work.  
The $u$ band shift is important for this paper, as it implies that
the stars are hotter than would be inferred from assuming that the SDSS
photometric zeropoints were perfectly on the AB system.  While the change is large, the primary reason
is known: the original zeropoints for the SDSS system laid out in 
\citet{Fuk96} neglected the effects of atmospheric extinction on the
bandpasses.  This is a 4\% shift in the $u$ band, 1\% in $g$, and
negligible elsewhere.  Hence, once this is corrected, the shifts in
$u$, $g$, $r$, and $i$ are all modest, less than 1.5\%, and do not 
affect our conclusions.  We will discuss our tests of the AB 
calibration in a later paper, but all tests to date recover a
shift in the $u$ band of about 4\%, i.e., the hypothesis of 
$u_{\rm AB}=u_{\rm SDSS}$ is not consistent with the data. 
Unless otherwise noted, all photometry is quoted on the SDSS system
without the AB corrections.  All analysis uses the AB corrected magnitudes.

\section{Hot DB White Dwarfs}
\subsection{Atmospheric Models}

For the spectroscopic fitting we used a grid of helium-rich model
spectra with 5 different traces of hydrogen: H/He = $10^{-20}$,
$10^{-5}$, $10^{-4}$, $10^{-3}$, $10^{-2}$. Effective temperatures ranged
from 10000 to 50000~K and logarithmic surface gravity ranged from 7.0 to
9.0. The numerical procedures of the atmospheric calculations and the
DK fitting routine (\S~\ref{sec:spect}) as well as the input physics are described in
past papers
\citep{Finley.Koester.ea97, Homeier.Koester.ea98,
Friedrich.Koester.ea00, Koester.Wolff00,
Koester.Napiwotzki.ea05}. 
We therefore give here only a brief
summary of those aspects that are important for hot, helium rich white
dwarfs and not covered in detail in the references. 

HeI bound-free cross sections were obtained from the Opacity Project
calculations as provided by the TOPBASE database at CDS
\citep{Seaton87, Cunto.Mendoza92, Cunto.Mendoza.ea93}. The Stark
broadening of the optical HeI lines uses the calculations of
\citet{bea97} with tables supplied to us through Dr.\ Thomas
Rauch (priv.\ comm.). One remaining uncertainty in DB models is the
treatment of the HeI resonance lines in the EUV. 
We have used the asymptotic wing formula described in \citet{Griem74}. Data for
electron and ion Stark broadening parameters were taken from that reference
and additionally from 
\citet{Dimitrijevic.Sahal-Brechot84, Dimitrijevic.Sahal-Brechot84*b,Dimitrijevic.Sahal-Brechot90}. 
In cool DB stars these resonance lines are very strong, violating the validity
criteria of the theory and raising conflicts with UV observations of white dwarfs
(the EUV resonance lines themselves have never been observed directly
in any white dwarf). We therefore have introduced artificial weakening of the
far wings beyond about 300\AA\ from the line center.
However, this uncertainty is not
relevant for the hot DBs considered here, since the resonance lines
are weaker than in the cool objects and do not overlap as strongly.

The optical HeI lines reach a maximum strength about 25,000~K.
This creates partial degeneracies in the spectral
fitting between models just below and above this point.  Above 30,000~K,
these degeneracies are milder, but the line strengths change rather gently
with temperature, leaving the determination of temperatures sensitive
to small systematic errors in the data and the models.  In the end, we
find that there are unexplained differences in the results at the
level of 1000~K.  This level of accuracy is sufficient for our purposes,
but the deviations suggest the opportunity for future improvements.

A major concern in the estimation of temperatures is the possible presence
of small levels of hydrogen in the atmospheres.  We assume that the hydrogen is
mixed homogeneously through the atmosphere.  The hydrogen alters the
line profiles of the helium lines so that a 100\% helium atmosphere model
would overestimate the temperature.  Similarly, the hydrogen slightly reddens the slope
of the spectral continuum, thereby causing a photometric temperature estimate to 
be overestimated.
The signature of hydrogen is the
appearance of Balmer lines, notably H$\alpha$, in the spectrum.  A H/He
fraction of $10^{-2}$ typically corresponds to a shift of about 4000~K
at these hot temperatures; however, most of the stars presented in this paper
do not show H$\alpha$ and H$\beta$ as strongly as this model would predict.
Hence, the possible temperature shifts are much smaller.
Throughout the paper, we use a H/He fraction of 
$10^{-2}$ for the 4 stars that show Balmer lines and $10^{-5}$ or $10^{-20}$ for the 
remainder of the sample,
the difference between the latter two being negligible.

\subsection{Selection}\label{sec:sel}

\begin{table*}[t]\footnotesize
\caption{\label{tab:basic}}
\begin{center}
{\sc Hot DB White Dwarf candidates \\}
\addtolength{\tabcolsep}{-2pt}
\begin{tabular}{clrrccccrrcrrcc} 
\doubleline
& & \multicolumn{3}{c}{Spectroscopy$^a$}
& \multicolumn{5}{c}{Imaging$^b$}
&
& \multicolumn{4}{c}{Astrometry$^d$} \\
Name & Class &
\multicolumn{3}{c}{Plate/Fiber/MJD} &
\multicolumn{5}{c}{Run/Rerun/Col/Field/Id} &
HS?$^c$ & 
\multicolumn{1}{c}{RA} &
\multicolumn{1}{c}{Dec} &
$\mu$$^e$ & Angle$^f$ \\ 
\doubleline
\input{db_table1.dat}
\doubleline
\end{tabular}
\end{center}
NOTES.---%
$^a$The identification numbers for the SDSS spectroscopy. \\
$^b$The identification numbers for the SDSS imaging. \\
$^c$Whether or not (Y or N) the object was targetted as a HOT\_STANDARD.
If so, the object received high priority in fiber allocation;
if not, the object was targeted through the star or serendipity sets
and recieved lower priority. \\
$^d$Proper motion from USNO-B catalog. \\
$^e$Proper motion in arcseconds per century.  The USNO catalog errors vary
but are of order unity in these units.\\
$^f$Direction of proper motion (North through East).
\end{table*}

\begin{table*}[t]\footnotesize
\caption{\label{tab:photom}}
\begin{center}
{\sc Photometry \\}
\begin{tabular}{ccccccc} 
\doubleline
Name & 
$u$ & $g$ & $r$ & $i$ & $z$ & $A_g$ \\
\doubleline
\input{db_table2.dat}
\doubleline
\end{tabular}
\end{center}
NOTES.---%
The SDSS photometry in the $u$, $g$, $r$, $i$, and $z$ bands, 
uncorrected for reddening or AB zeropoints.
The error in the last two digits are given in parentheses.  This error
is only the photon Poisson error; it does not include field-to-field calibration
errors, which is estimated at 0.01 mag in $g$, $r$, $i$, 0.02 mag in $z$, and 
0.03 mag in $u$ \protect\citep{Ive04}.
An asterisk indicates that the photometry in that band was flagged during
processing as suspect \protect\citep{Kle04}; 
we have excluded these bands from our model fitting.
The \protect\citet{SFD} extinction prediction for the $g$ band
is given in the last column.
\end{table*}

We draw our sample from the SDSS DR4 white dwarf catalog \citep{Eis05}.
The parent set for this catalog was selected by their blue colors and
stellar redshifts.  The stars were then fit to stellar atmospheres
by the autofit method described in \citet{Kle04} and \citet{Eis05}.
Those with temperature and surface gravities indicative of white dwarfs
were denoted as such.  Stars with unusual results were flagged for visual
inspection and could be added to or removed from the white dwarf list.

In brief, the autofit method performs a $\chi^2$ minimization of the
spectrum and photometry relative to the grid of atmospheric models to
yield an estimate of the temperature and surface gravity of the star.
Two grids of atmospheres are used, pure hydrogen and pure helium;
typically one model grid has a signficantly lower best $\chi^2$ than 
the other.
All model spectra are convolved to the spectral resolution of the data.
Masked pixels are excluded, as are any photometric bands with suspicious
warning flags.  Errors on the fitted parameters are computed by taking
moments of the likelihood function $\exp(-\chi^2/2)$.  The continuum is
controlled by marginalizing over an unknown polynomial of order 5 that
multiplies the model spectrum.  This order of polynomial is sufficiently
general to be completely degenerate with the reddening or with any
effect from atmospheric dispersion moving the star relative to the SDSS
fiber entrance.  The spectra are fit over the range 3900\AA\
to 6800\AA.
Our 5th order polynomial will remove
errors in the SDSS spectrophotometry on scales above 500\AA, but errors
on smaller scales would be passed through and could affect the fits if
ill-placed.  However, the SDSS fluxing does appear to be good, at least
in the mean, on these smaller scales, save for a 5\% bump at 3930\AA.

From the DR4 catalog, we select all white dwarfs, regardless of
classification, that have dereddened $g<19.5$ and that were fit best by a
helium atmosphere hotter than 28,000~K.  We repeated the autofit analysis
on these stars, extending the helium grid up to 50,000~K.  We used maximum
wavelengths of 6800\AA\ and 5400\AA\ 
and also used a model grid with 1\% hydrogen.
We then inspected the stars, studying the spectra and photometry along with the 
various autofit results.
Some of the stars were subdwarfs; these changes in classification have been reflected
in the released DR4 catalog.  Others were DO stars that were hotter than
50,000~K, which we decided not to include in this paper as they are
much hotter than the putative DB gap.  One star was a cooler magnetic
DB whose spectrum had drastically biased the fit.  Three other stars
were rejected because closer inspection led us to conclude that they
were either considerably cooler or so noisy as to be inconclusive.

After the inspection, we are left with 28 stars that we believe to be hot DB or cool DO stars.
19 of these are from the Hot Standard class.
The basic information on these stars is
given in Table \ref{tab:basic}, including the USNO-B proper
motion.  The SDSS photometry is given in Table \ref{tab:photom}.
Some bands have photometric flags set that lead us to be suspicious
of the photometry; these bands have been marked with asterisks and 
excluded from the autofit and subsequent analysis.

\subsection{Follow-up Data}

\begin{figure*}[p]
{\epsfysize=1.4in\epsffile{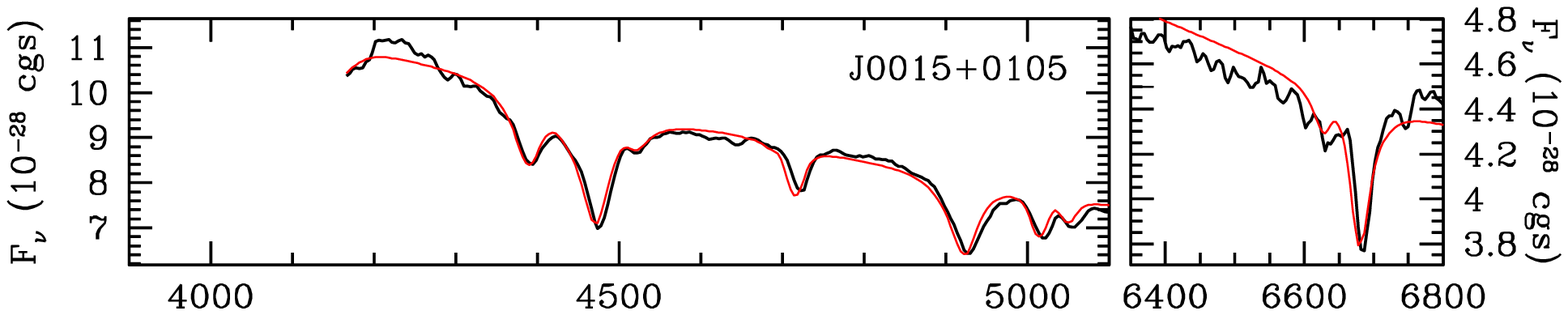}}\vspace{-8pt}
{\epsfysize=1.4in\epsffile{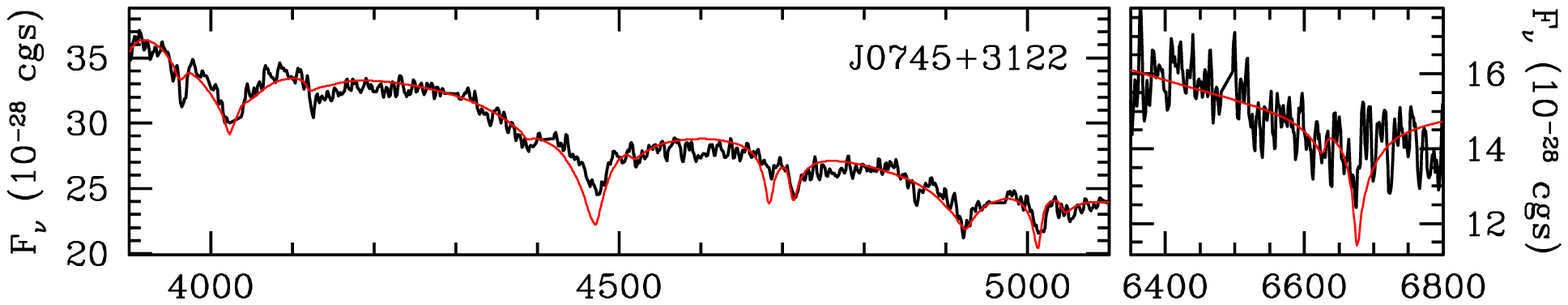}}\vspace{-8pt}
{\epsfysize=1.4in\epsffile{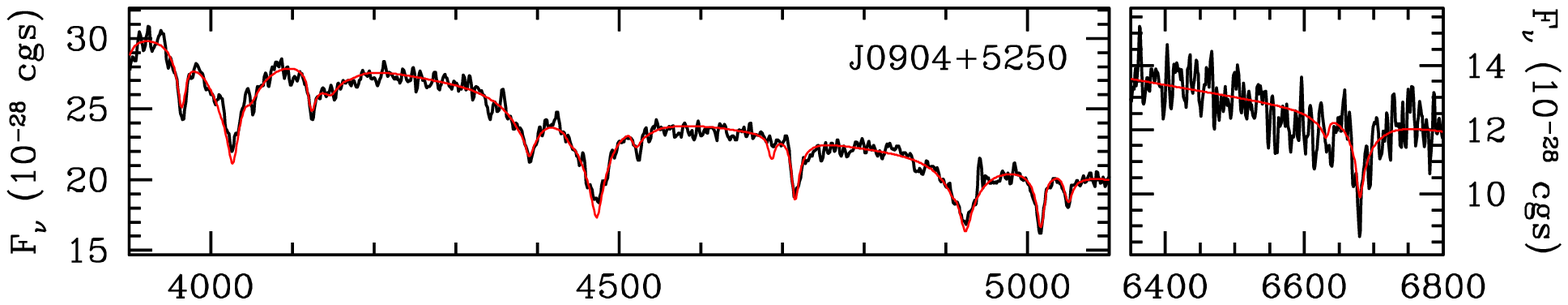}}\vspace{-8pt}
{\epsfysize=1.4in\epsffile{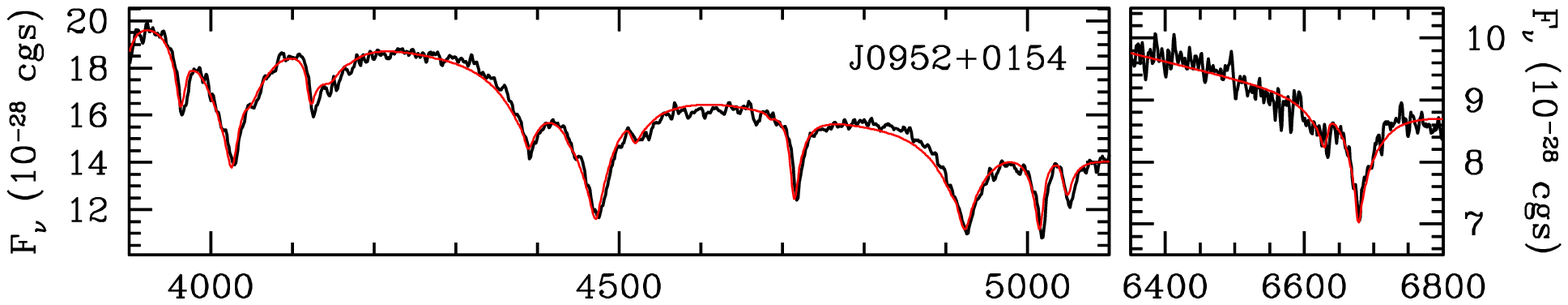}}\vspace{-8pt}
{\epsfysize=1.4in\epsffile{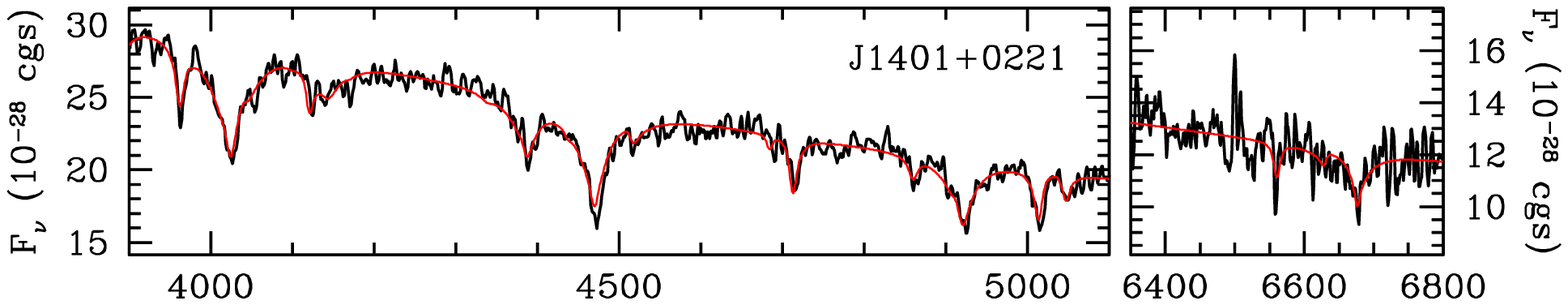}}\vspace{-8pt}
{\epsfysize=1.4in\epsffile{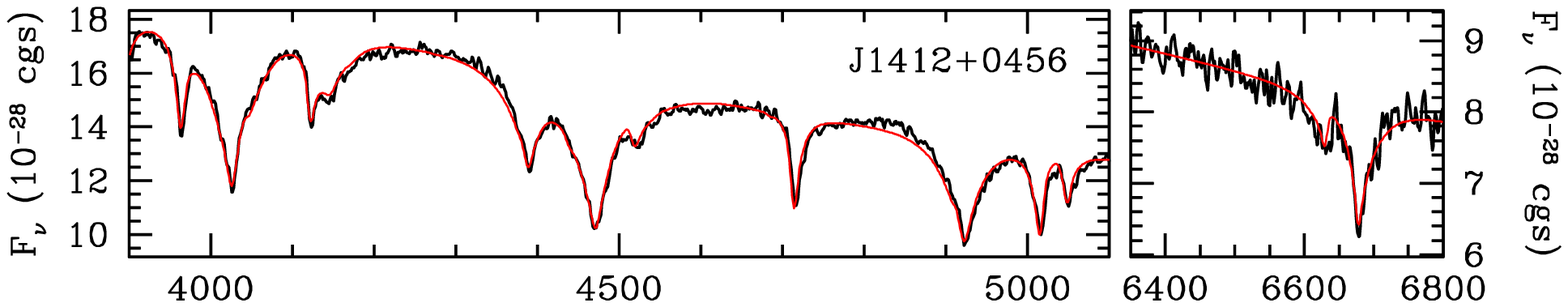}}\vspace{-8pt}
\caption{\label{fig:mmt}
Spectra from the MMT overlaid with the best-fit model.
The model has been refluxed with 6 polynomials to match the observed spectrum.
The spikes are a few surviving cosmic rays.
}
\end{figure*}

\begin{figure*}[p]
{\epsfysize=1.4in\epsffile{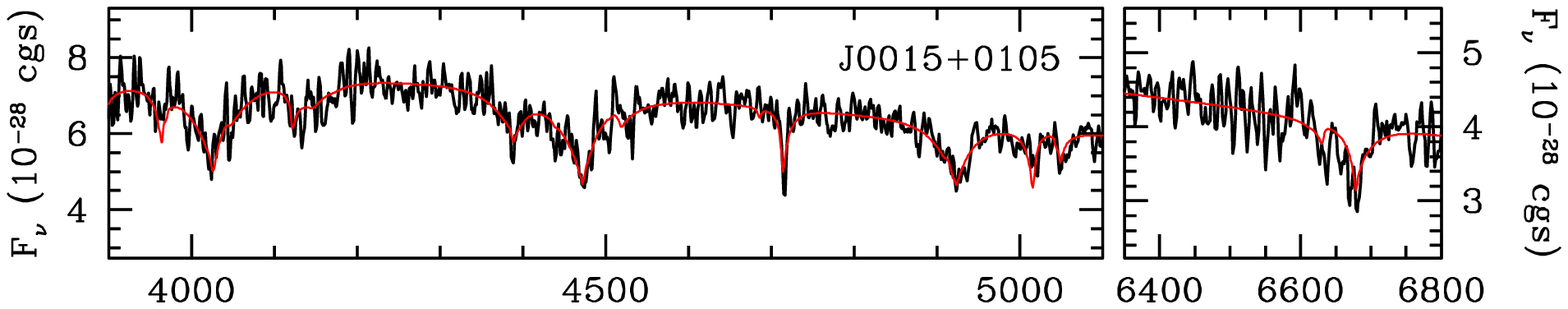}}\vspace{-8pt}
{\epsfysize=1.4in\epsffile{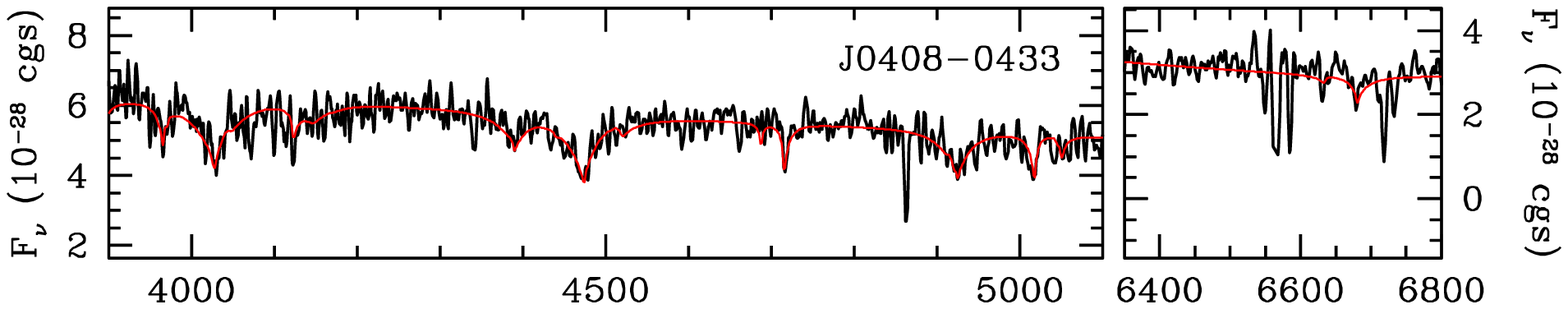}}\vspace{-8pt}
{\epsfysize=1.4in\epsffile{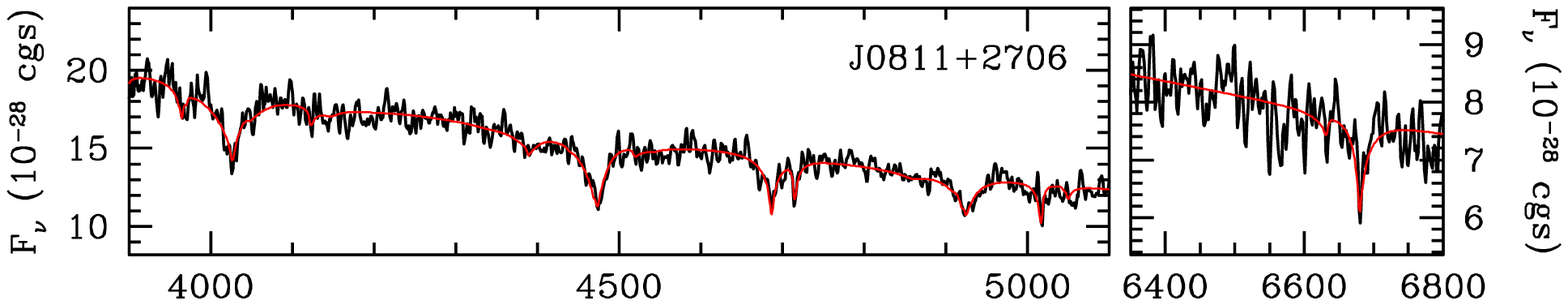}}\vspace{-8pt}
{\epsfysize=1.4in\epsffile{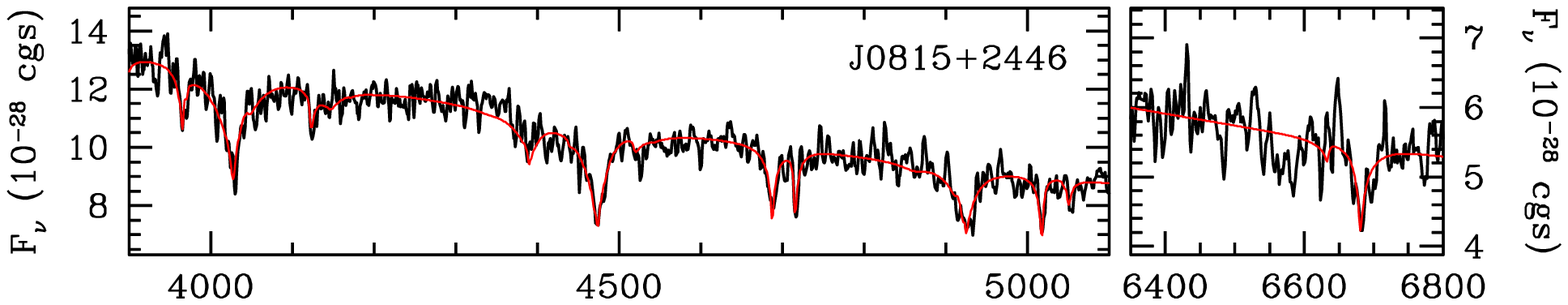}}\vspace{-8pt}
{\epsfysize=1.4in\epsffile{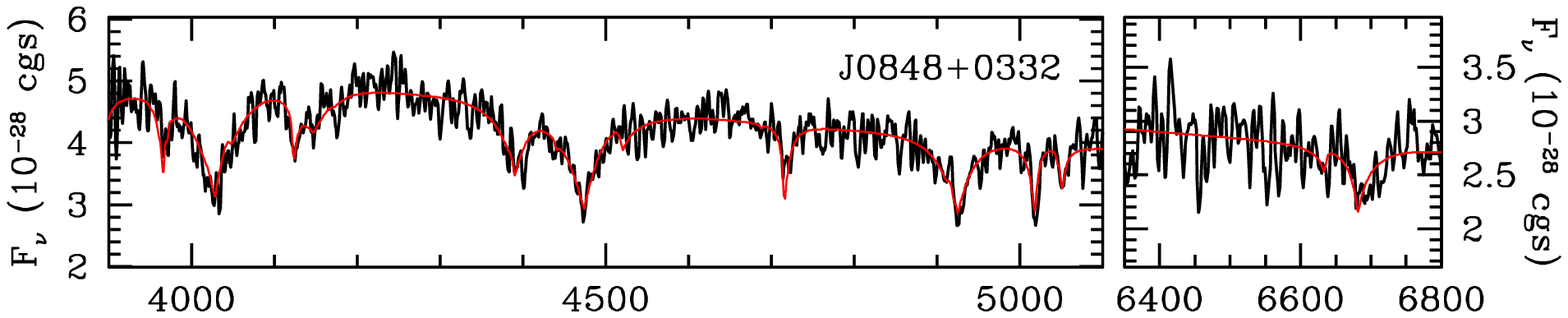}}\vspace{-8pt}
{\epsfysize=1.4in\epsffile{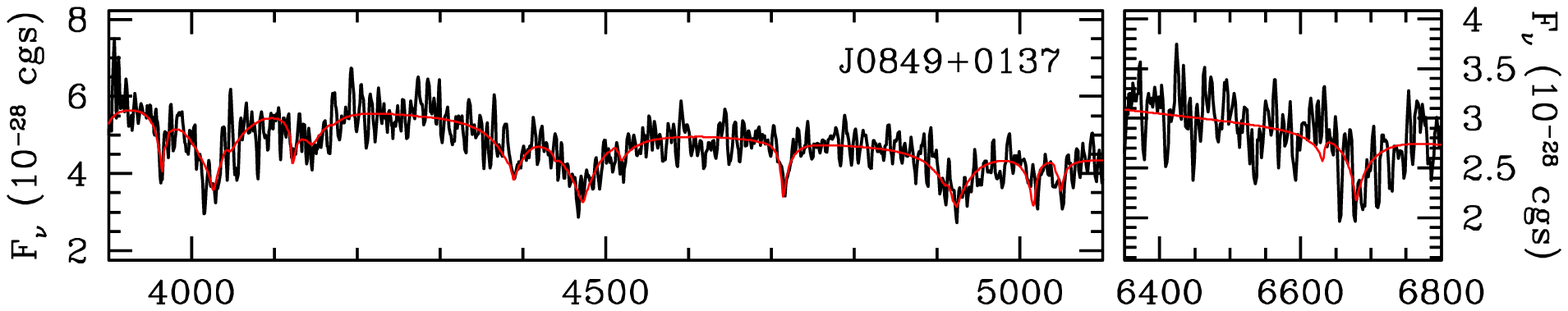}}\vspace{-8pt}
\caption{\label{fig:sdss1}
Spectra from the SDSS overlaid with the best-fit model.
The model has been refluxed with 6 polynomials to match the observed spectrum.
}
\end{figure*}

\begin{figure*}[p]
{\epsfysize=1.4in\epsffile{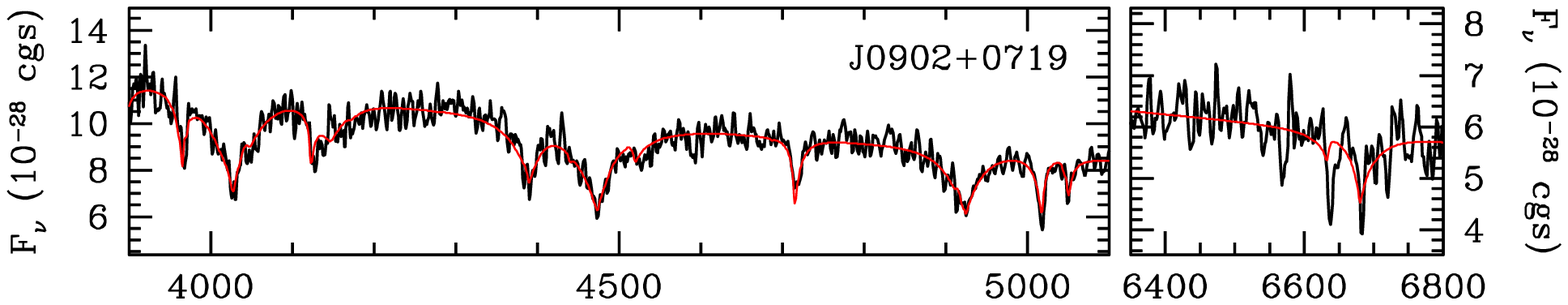}}\vspace{-8pt}
{\epsfysize=1.4in\epsffile{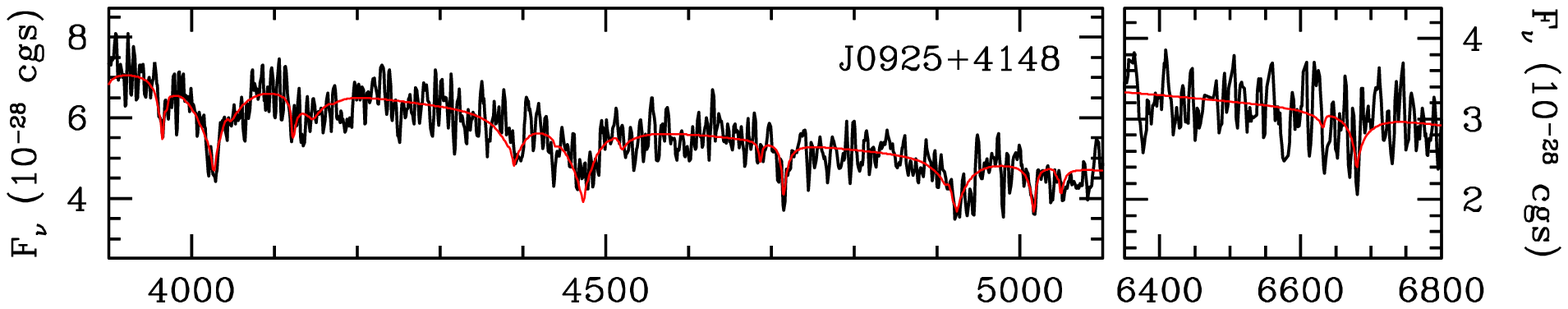}}\vspace{-8pt}
{\epsfysize=1.4in\epsffile{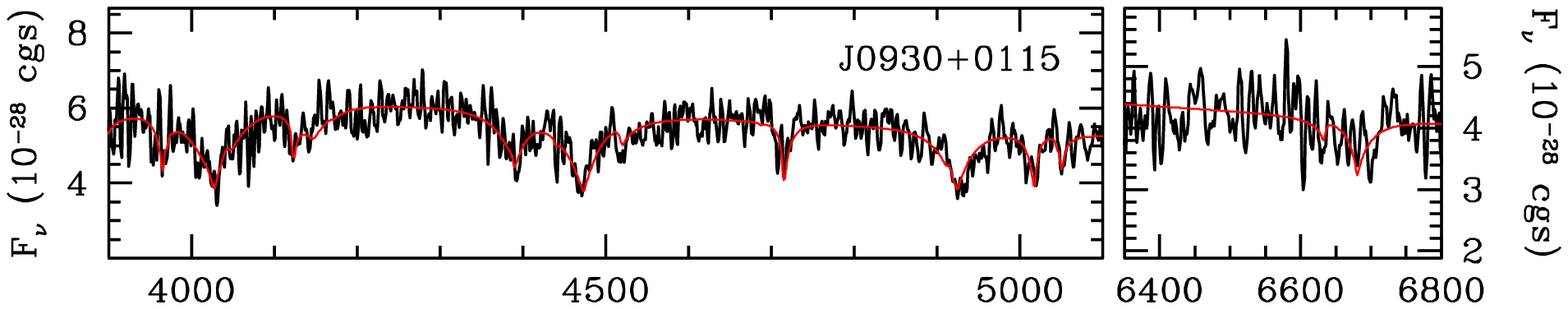}}\vspace{-8pt}
{\epsfysize=1.4in\epsffile{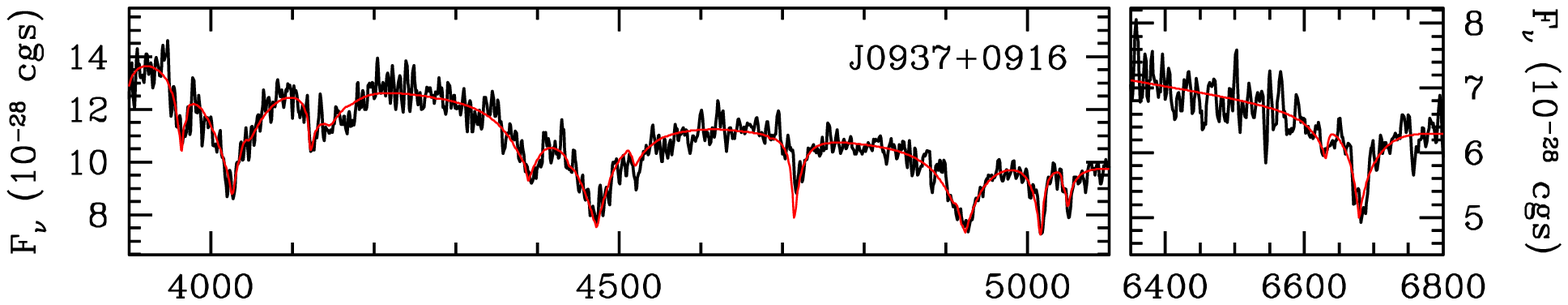}}\vspace{-8pt}
{\epsfysize=1.4in\epsffile{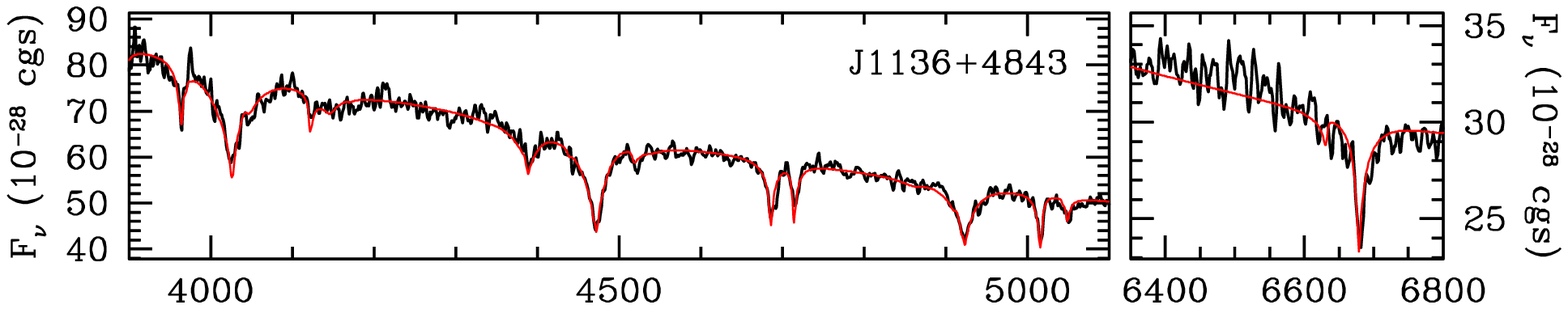}}\vspace{-8pt}
{\epsfysize=1.4in\epsffile{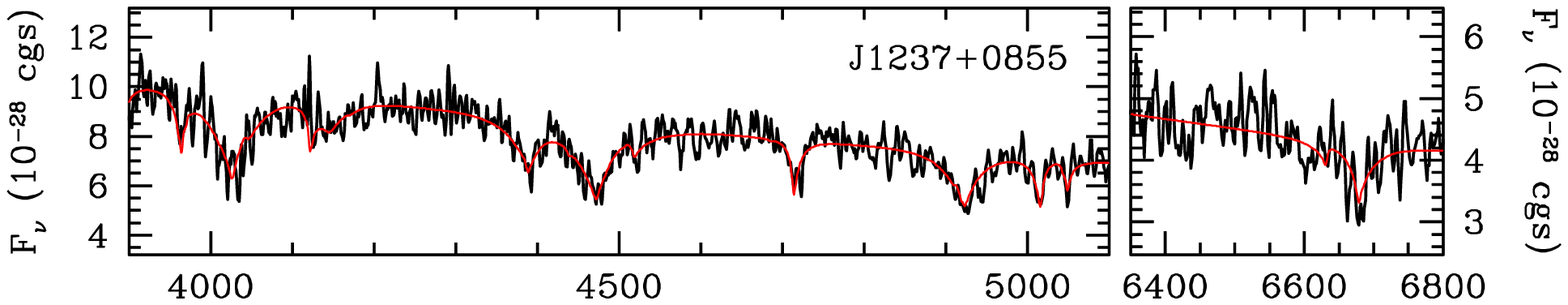}}\vspace{-8pt}
\caption{\label{fig:sdss2}
More spectra from the SDSS.
}
\end{figure*}

\begin{figure*}[p]
{\epsfysize=1.4in\epsffile{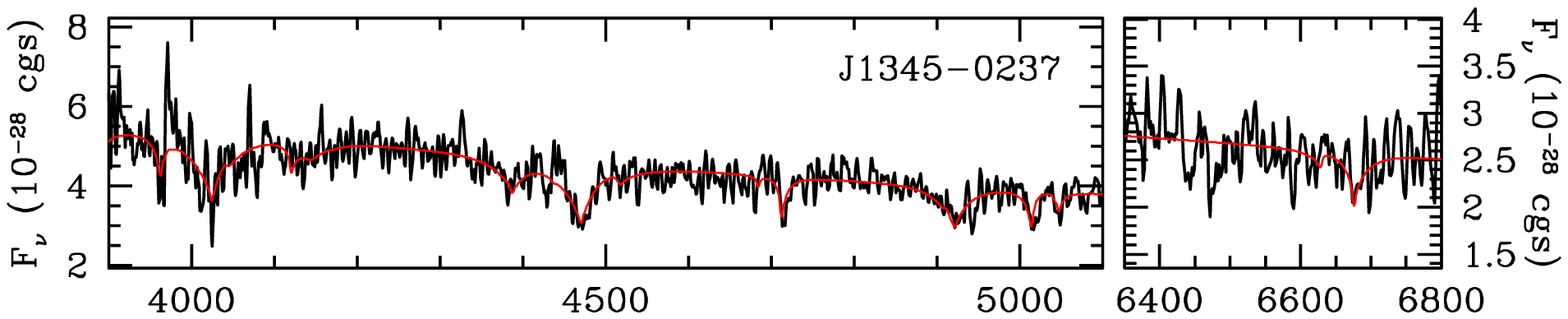}}\vspace{-8pt}
{\epsfysize=1.4in\epsffile{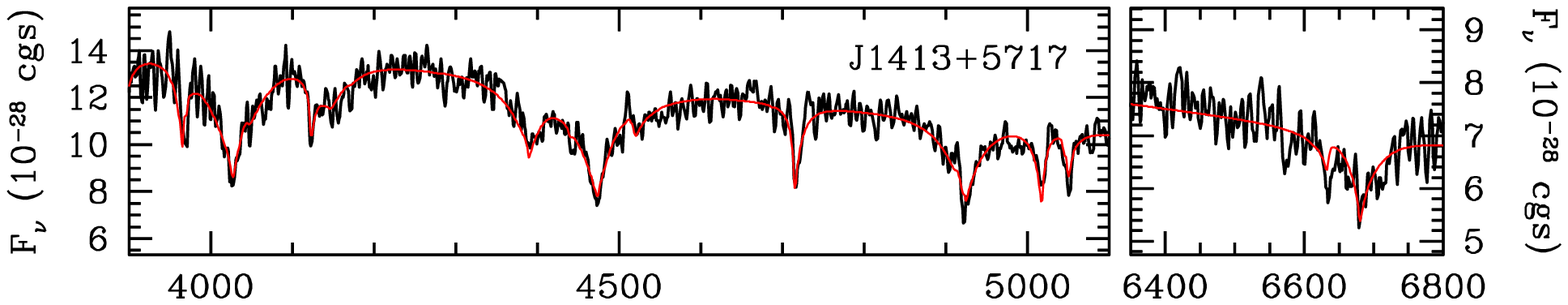}}\vspace{-8pt}
{\epsfysize=1.4in\epsffile{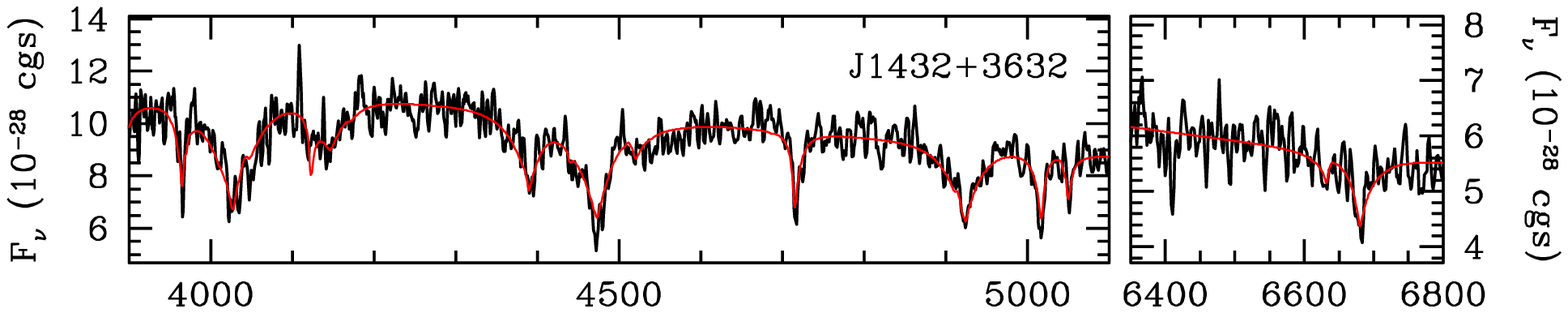}}\vspace{-8pt}
{\epsfysize=1.4in\epsffile{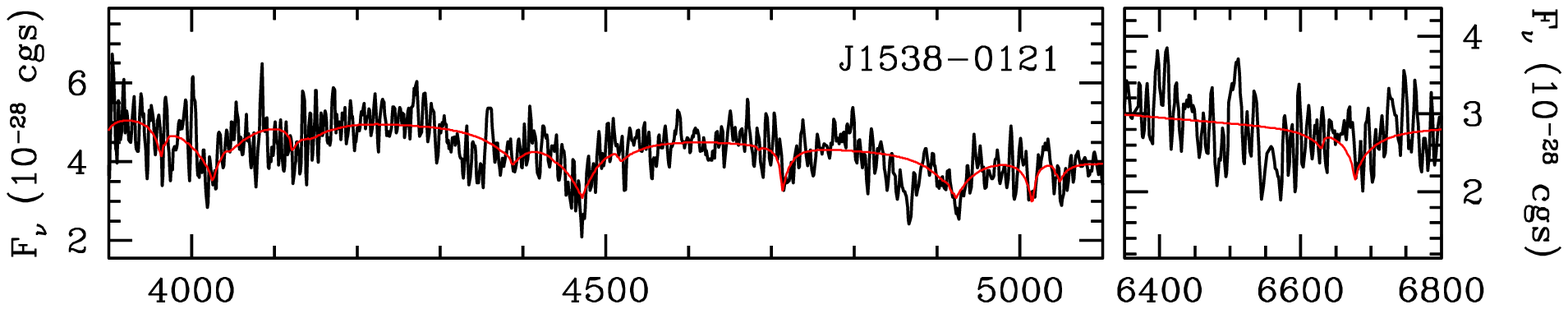}}\vspace{-8pt}
{\epsfysize=1.4in\epsffile{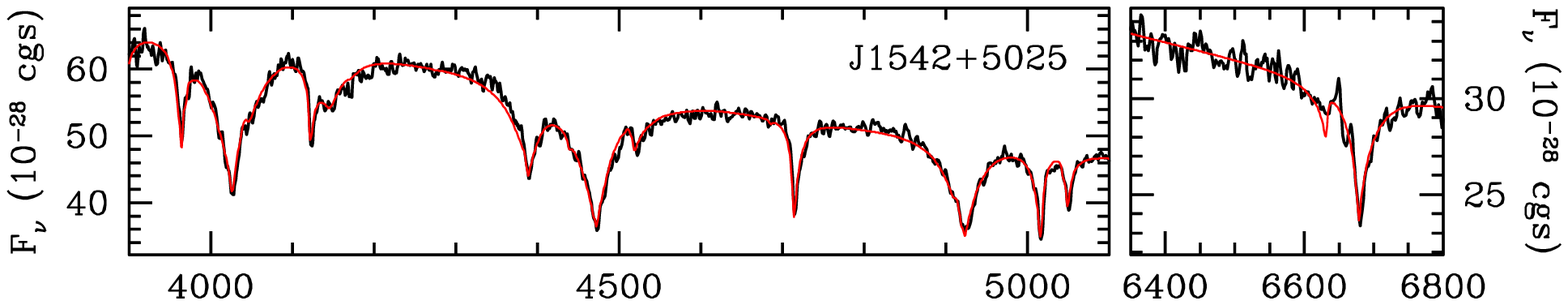}}\vspace{-8pt}
{\epsfysize=1.4in\epsffile{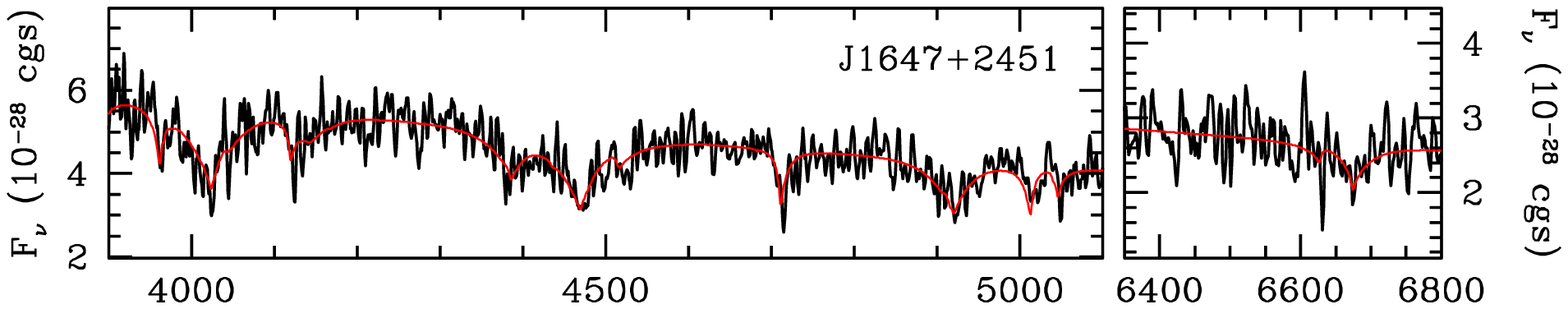}}\vspace{-8pt}
\caption{\label{fig:sdss3}
More spectra from the SDSS.
}
\end{figure*}

\begin{figure*}[p]
{\epsfysize=1.4in\epsffile{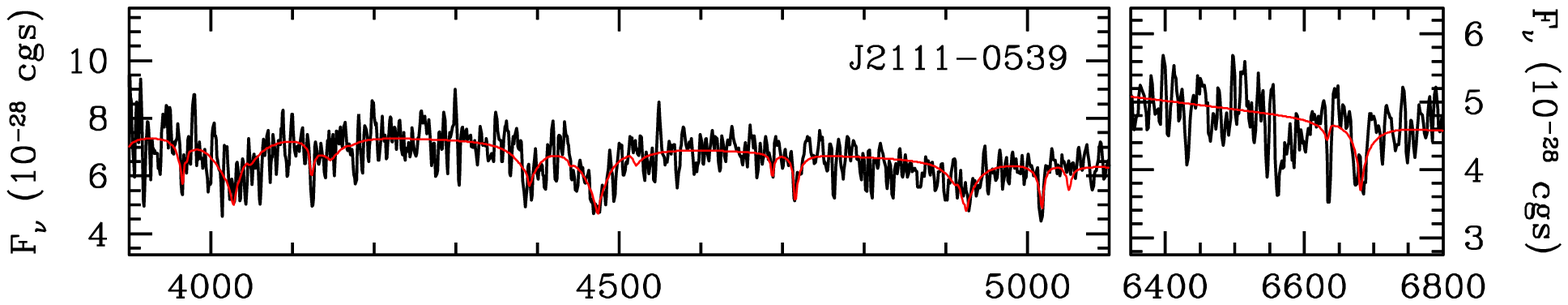}}\vspace{-8pt}
{\epsfysize=1.4in\epsffile{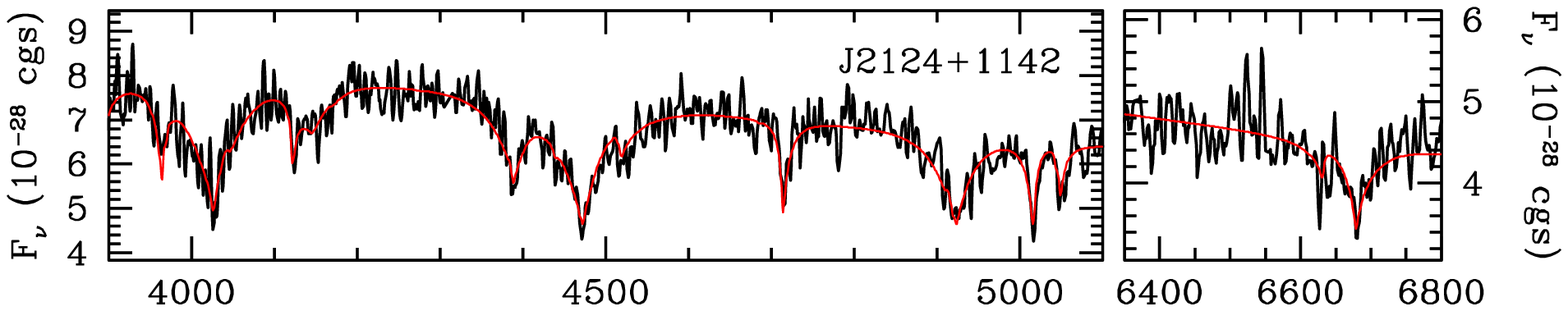}}\vspace{-8pt}
{\epsfysize=1.4in\epsffile{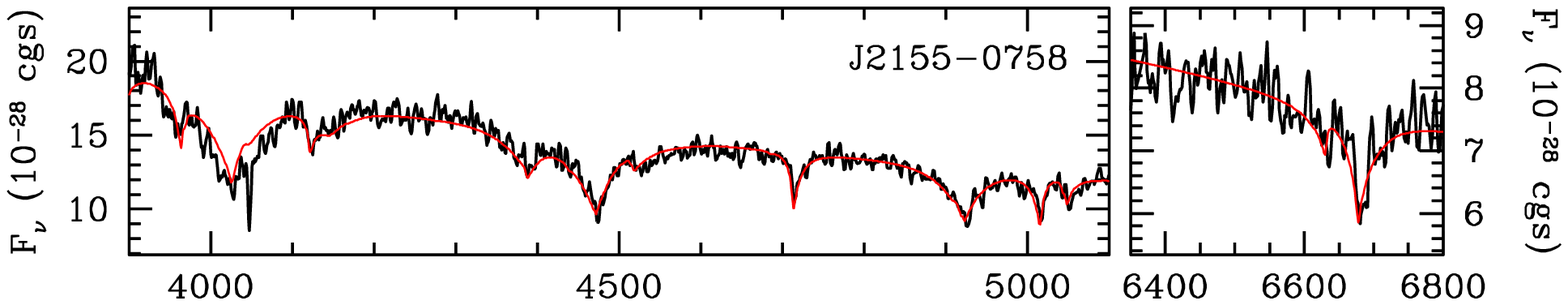}}\vspace{-8pt}
{\epsfysize=1.4in\epsffile{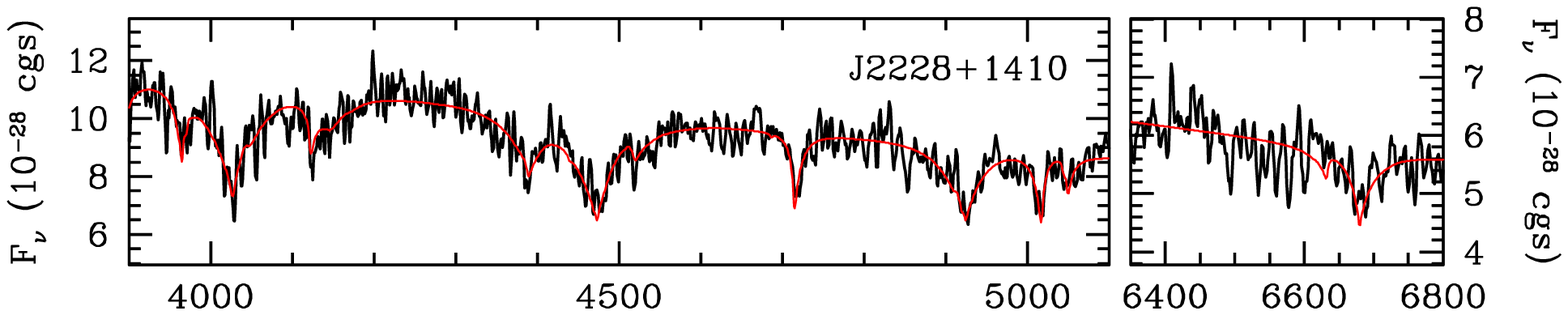}}\vspace{-8pt}
{\epsfysize=1.4in\epsffile{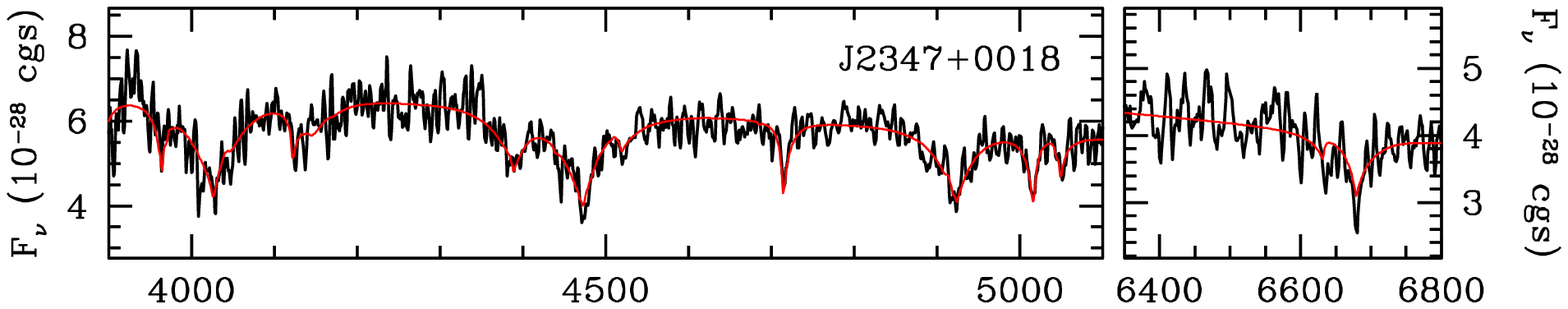}}\vspace{-8pt}
\caption{\label{fig:sdss4}
More spectra from the SDSS.
}
\end{figure*}

Unfortunately, the signal-to-noise ratio in the red portion of the
SDSS spectra was not always sufficient to detect weak H$\alpha$
lines in these very blue stars.  We therefore acquired higher
signal-to-noise ratio spectra at the 6.5-meter MMT in order to refine the
temperature estimates and search for the possible Balmer lines.
These spectra are shown in Figure \ref{fig:mmt}.  The SDSS
spectra for stars not observed at the MMT are shown in 
Figures \ref{fig:sdss1}--\ref{fig:sdss4}.

One star (J0015+0105) was observed with the SpecPol spectrometer
\citep{Sch92,Smi03} on October 29, 2003.
The spectral resolution is
low, about 18\AA\ FWHM through the $1''.5$ slit, 
but the throughput is very high, permitting
a good constraint on the Balmer lines in
the 45 minute exposure.  The flux calibration was 
performed using the standard star G~191-B2B.

Five stars 
(J0952+0154, J1401+0221, J0904+5250, J1426+0456, J0745+3122)
were observed with the Blue Channel spectrograph on the 
night of January 19, 2004.  Seeing conditions were poor, such that
we used a wide slit ($1''.5$--$2''$).  A 500 mm$^{-1}$ grating was used, resulting
in roughly 6\AA\ FWHM spectral resolution over a wavelength range
of 3900\AA\ to 6800\AA.  The spectra were reduced with standard
IRAF packages and fluxed using the spectrophotometric standard 
stars G~191-B2B and Feige~66.  Exposure times ranged between 20 and 45
minutes per star.

In addition to our 5 targets, we observed the well-studied, hot,
pulsating DBV star GD 358 with the intent of testing our modeling.  We find,
however, that this star is sufficiently close to the peak strengths of
the He I lines (near 25,000~K) that its
temperature fits are unstable in a manner that is not expected to
extend to higher temperatures.  
When we fit GD 358 with autofit over the
full spectral range, we obtain a temperature of 27,700~K, but when we
restrict to blueward of 5400\AA\ (where most but not all of the strong
lines are), the derived temperature is 23,900~K.  
Making a similar restriction on our
primary stars typically changes our fits by only a few hundred degrees, 
and only one star by more than 1000~K (SDSS J1538--0121, which has other
discrepencies to be described in \S~\ref{sec:stars}) 
The DK fitting package has similar trouble with GD 358.

\subsection{Photometric Temperature Estimates}

\begin{table*}[t]\footnotesize
\caption{\label{tab:photo_fit}}
\begin{center}
{\sc Photometric Fits \\}
\begin{tabular}{ccccccc} 
\doubleline
Name & 
$u-g$ (AB) & $g-i$ (AB) & $T_{photo}$ & 95\% range & $\chi^2$ & DOF$^a$ \\
\doubleline
\input{db_table3.dat}
\doubleline
\end{tabular}
\end{center}
NOTES.---%
The colors include a maximal reddening correction from \protect\citet{SFD}
and a 4\% and 1.5\% AB correction for $u-g$ and $g-i$, respectively,
making the colors bluer.
The best-fit temperature for a pure Helium atmosphere with $\log(g)=8.0$ 
is listed, using only the photometry and the errors (suspect bands are
marked by asterisks and given no weight).  The 95\% confidence
interval is listed; strictly speaking, these are simply the $\Delta\chi^2=4$
values.  We omit the upper bound if it exceeds 50,000~K, the maximum of
our grid.  All temperatures are in 1000's of Kelvin.  A maximal
reddening correction is assumed.  The $\chi^2$ of the best fit is given.\\
$^a$Degrees of freedom available to fit. \\
$^b$The colors and fit are based on the stacked photometry listed in 
\S\protect\ref{sec:stars}
rather than single-epoch photometry in Table \protect\ref{tab:photom}. \\
$^c$The fit is to a model with 1\% hydrogen fraction rather than pure helium. \\
$^d$The M star companion is bright enough that we have excluded the 
$i$ and $z$ bands from the fit.
\end{table*}

At these temperatures, the predicted photometric colors depend
sensitively on temperature but are insensitive to surface gravity.  At
30,000~K, the $u-g$ and $g-i$ colors change by $-0.014$ mag and $-0.011$
mag per 1000~K, respectively, but a surface gravity change of a
factor of 10 (i.e., 1 dex) changes these colors by only 0.023 mag and 0.024 mag,
respectively.  We therefore perform a fit of the model predictions
directly to the photometry alone, assuming that the stars have pure
helium atmospheres and $\logg=8.0$.  The surface gravities of typical
white dwarfs and DBs in particular are known to be clustered around
$\logg\approx8$ \citep{bea95}, and the dispersion around that point is small enough as
account for well less than 1000~K of uncertainty.

We present these photometry-only fits in Table \ref{tab:photo_fit}.
All stars were dereddened by the full amount of the \citet{SFD} prediction.
A 30,000K DB white dwarf with $\log g=8.0$ would have $M_g\approx10$
(Bergeron, private communication).
Our fainter stars, with $g>18$, are therefore at
least 400 parsecs away, and since all of the targets are at high Galactic
latitude, the reddening assumption is appropriate for the fainter stars.
However, the brightest stars may be over-corrected, which would
cause an overestimate of their temperature.
We further apply a correction to move the SDSS photometry to the 
AB system, as discussed in \S~\ref{sec:sdss}.  
For reference, the 30,000 K DB model has synthetic AB colors of
$u-g=-0.267$ and $g-i=-0.770$, i.e., SDSS colors of $u-g=-0.227$ and $g-i=-0.755$.  
Finally, we add calibration uncertainties in quadrature to the errors
in all bands; these are taken to be 0.01 mag in $g$, $r$, and $i$, 
0.02 mag in $z$, and 0.03 mag in $u$ \citep{Ive04}.

Two of the candidates (J0015+0105 and J2347+0018) 
lie on the southern Galactic cap equatorial strip,
where the SDSS has imaged repeatedly.  Therefore these objects
have extremely precise photometry.  In both cases, we use 11 epochs
of photometry.  These 11 agree with each other and are consistent with
Table \ref{tab:photom}.  
The weighted averages are
$u=18.799$, $g=18.949$, $r=19.378$, $i=19.653$, and $z=19.898$
for SDSS J234709.3$+$001858
and
$u=18.683$, $g=18.945$, $r=19.393$, $i=19.708$, and $z=20.017$ for 
SDSS J001529.7$+$010521.
In both cases, the 
errors derived from the variance of the 11 epochs
are 0.015 mag or smaller for all bands, save for 0.04 mag in $z$.
Because some but not all of the calibration uncertainties will be decreased 
in multiple observations, we adopt 0.015 mag (0.04 mag in $z$) as the errors.

We regard the photometric fits as a robust temperature estimate.
The precision of the temperature estimate is typically limited by
the photometric accuracy, particularly in the $u$ band.  This is 
usually the result of the statistical errors, but a floor of about
$\sigma=1000$~K is set by the knowledge of the photometric zeropoint 
and uncertainties in the reddening correction.  Typically, the errors
are closer to 2000~K.  Nevertheless, this constrains many of the stars
to be hotter than 30,000 K, although some are better fit in the 28,000 K
range.

\subsection{Spectroscopic Temperature Estimates}
\label{sec:spect}

\newcommand{\nommt}{& \nodata & \nodata}
\begin{table*}[t]\footnotesize
\caption{\label{tab:spect_fit}}
\begin{center}
{\sc Spectroscopic Fits \\}
\begin{tabular}{clcllcllc} 
\doubleline
& & & \multicolumn{3}{c}{DJE Fitting} 
& \multicolumn{3}{c}{DK Fitting} \\
Name & Class & $(S/N)_g$ & 
$T_{sp}$ & $\log g$ & $\chi^2$ & $T_{sp}$ & $\log g$ & $\chi^2$ \\
\doubleline
\input{db_table4.dat}
\doubleline
\end{tabular}
\end{center}
NOTES.---%
The MMT lines refer to the MMT spectrum of the star immediately above.
For the SDSS spectra, the formal errors on the last two digits are given in parentheses.
However, the discrepencies between the two fitting packages suggest 
that the systematic errors are important as well, particularly
for the surface gravities.
The signal-to-noise ratio is quoted for the $g$ band.  We have omitted
quotes of the signal-to-noise ratio and formal errors from the MMT data
because these reductions are not as well characterized as the SDSS.
Of course, the S/N ratio is considerably higher.\\
$^\dagger$Using a H/He ratio of 1\%, which is a representative value but not
optimized.  
The SDSS spectrum of J1401+0221 fits to 38,900K and $\log g=8.01$ with a pure He atmosphere.\\
\end{table*}

We fit the SDSS and MMT spectra to the atmosphere models by two different
methods.  These two packages share the same grid of atmosphere models and
the same spectral reductions but are otherwise independent.  

The first is the autofit method described in \S~\ref{sec:sel}.
These resulting temperatures and gravities are presented in Table 
\ref{tab:spect_fit} as the DJE numbers.  There is a mild covariance between temperature
and gravity, such that higher gravities imply higher temperatures.
Note that autofit does include the photometric data as well, but the 
photometry is much less constraining (at least in a statistical sense)
than the spectra and so it has relatively little pull on the results.

The second model fitting was performed by Detlev Koester and is listed in Table
\ref{tab:spect_fit} as the DK numbers.  In this method, the model is fit
to the observed spectrum in a number of preselected continuum regions
between the spectral lines.  The flux correction factors determined
for these regions are then quadratically interpolated for the whole
spectrum. After this normalization, the $\chi^2$ statistic is determined
only in the regions containing the lines. In essence, this means that
only the information contained in the line shape relative to the adjacent
continuum is used.  The photometry is not used as a constraint.

\begin{figure}[p]
\centerline{\epsfxsize=5in\epsffile{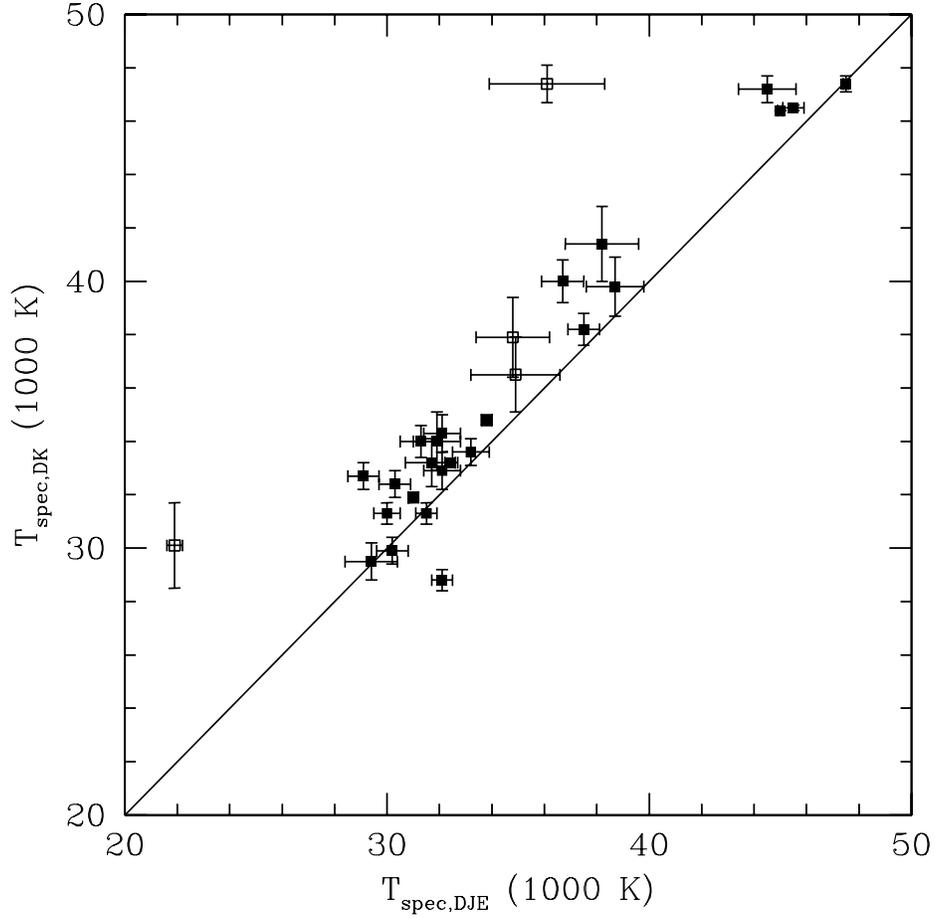}}
\caption{\label{fig:Tspec_Tk}%
The comparison of the temperatures inferred from the two different
spectroscopic fitting methods for the SDSS spectra.  1 $\sigma$ error bars
are shown.  The diagonal line is equality.  Solid symbols show
objects fit to pure helium atmospheres; hollow symbols show
the objects fit to 1\% hydrogen models.  Overall the correlation is 
good.  Removing the three outliers, the Koester fits are about 1500~K
hotter than the autofit results, with a scatter of about 1000~K.  This
gives our estimate of the systematic errors in the fitting.
}
\end{figure}

Figure \ref{fig:Tspec_Tk} compares the results from the two
fitting methods.  We are generally encouraged by the agreement.  All of
the fits to MMT data agree well, to better than 700~K.  For the SDSS spectra,
there are a few outliers (to be discussed later), but most fits agree.
There does appear to be a trend for the DK temperatures to be higher
than the DJE temperatures by about 1500~K.  This difference seems to be
smaller at higher signal-to-noise ratio, only 800~K at $S/N>12$ per pixel
and better for the MMT data, and increases for noisier data, suggesting
that it is related to some difference in how pixels are weighted.
The surface gravities agree to reasonable accuracy, with a mean offset
of 0.05 dex.  Removing these offsets, the scatter between the two fits is about 1000~K and about
0.2 dex in $\logg$ overall and about 25\% better in the higher signal-to-noise ratio
data.  We take this to be an indication of the systematic
uncertainties in the fitting.

\begin{figure}[p]
\centerline{\epsfxsize=5in\epsffile{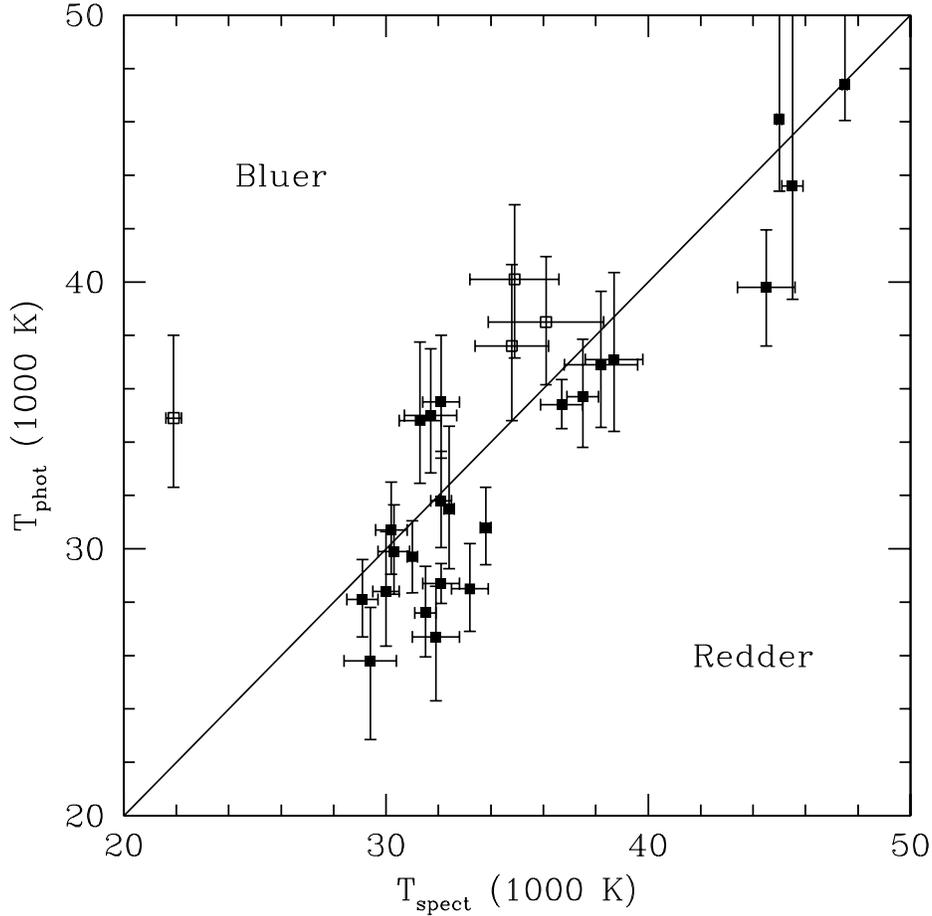}}
\caption{\label{fig:Tspec_Tphot}%
The correlation of the temperature inferred from the photometry
with that inferred from the spectroscopy.  1 $\sigma$ error bars
are shown.  The diagonal line is equality.  Solid symbols show
objects fit to pure helium atmospheres; hollow symbols show
the objects fit to 1\% hydrogen models.  The words ``Bluer''
and ``Redder'' are to remind the reader that when the photometric
temperature is hotter than the spectroscopic temperature, then
the colors are bluer than the spectroscopic best-fit model would
predict, and vice versa.  Overall the correlation is good.  There
may be a tendency at the lowest temperatures for the photometric
temperature to be lower than the spectroscopic temperature.  On 
the other hand, we required $T_{\rm spect}>28,000$~K to enter the
sample, so this may be a selection effect.
}
\end{figure}

\begin{figure}[p]
\centerline{\epsfxsize=5in\epsffile{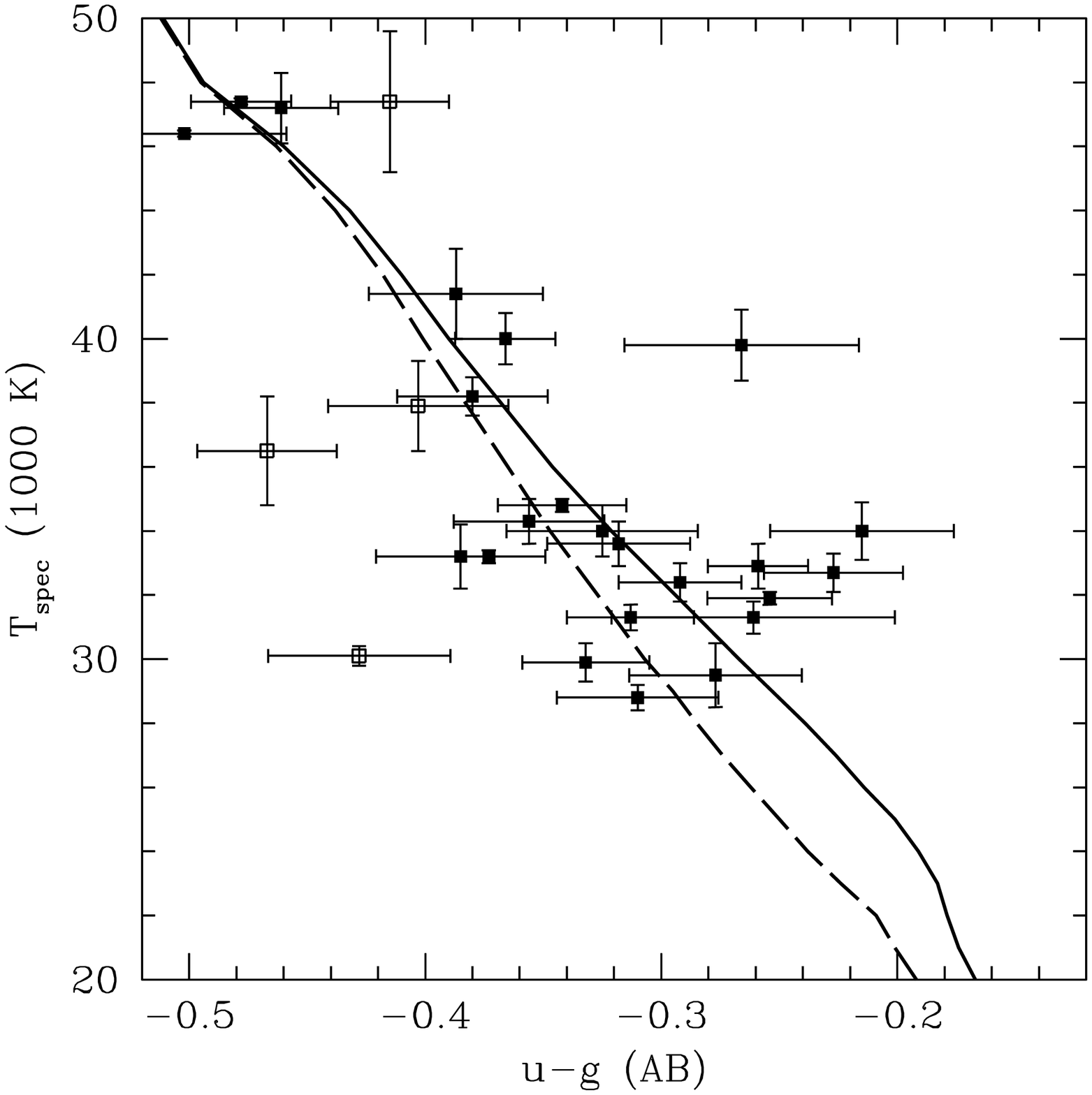}}
\caption{\label{fig:Tspec_ug}%
The correlation of the temperature inferred from the spectroscopy 
with the $u-g$ color.  The $u-g$ color has been adjusted to AB
zeropoints and maximally dereddened; both of these corrections
make the color bluer.  Error bars are 1~$\sigma$.  Solid symbols show
objects fit to pure helium atmospheres; hollow symbols show
the objects fit to 1\% hydrogen models.  Stars with 
bad quality flags for the $u$ or $g$ photometry have been omitted.
The solid line is the temperature-color relation predicted for a pure helium
atmosphere; the dashed line is that for a 1\% hydrogen contamination.
} 
\end{figure}

\begin{figure}[p]
\centerline{\epsfxsize=5in\epsffile{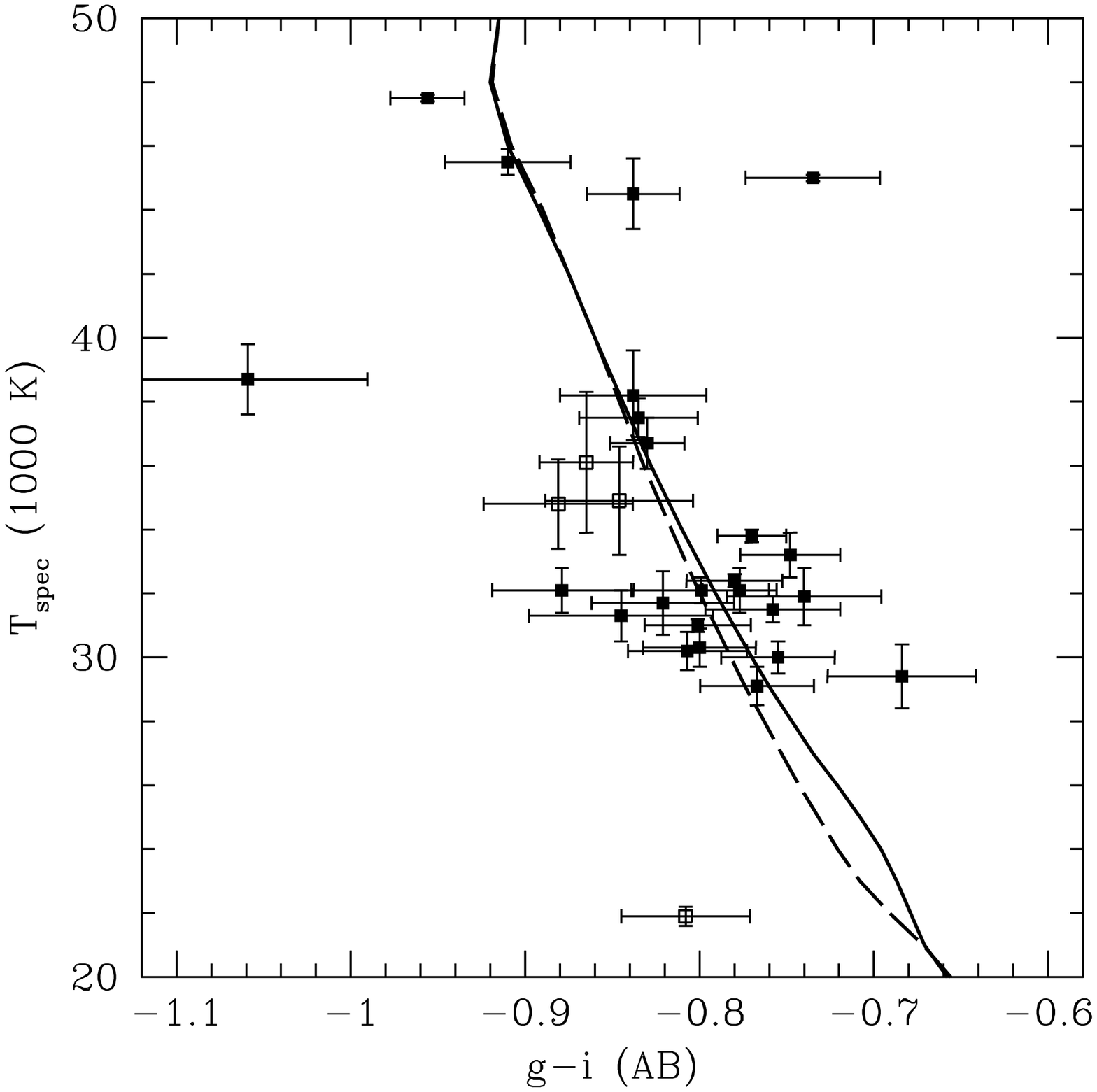}}
\caption{\label{fig:Tspec_gi}%
The correlation of the temperature inferred from the spectroscopy 
with the $g-i$ color.  The $g-i$ color has been adjusted to AB
zeropoints and maximally dereddened; both of these corrections
make the color bluer.  Error bars are 1~$\sigma$.  Solid symbols show
objects fit to pure helium atmospheres; hollow symbols show
the DBA objects fit to 1\% hydrogen models.  Stars with 
bad quality flags for the $g$ or $i$ photometry have been omitted.
The solid line is the temperature-color relation predicted for a pure helium
atmosphere; the dashed line is that for a 1\% hydrogen contamination.
}
\end{figure}

Figure \ref{fig:Tspec_Tphot} compares the results between the DJE fits 
and the photometric fits.  Again, the results agree well, save for one
outlier to be discussed later.  There is a hint that the photometric
temperatures are systematically slightly lower than the DJE spectral
fits (which in turn are slightly lower than the DK spectral fits).  In 
other words, the photometry is slightly redder than the models would
predict.  However, one can see from the Figure that this offset is 
at most 1000~K, which is comparable to the level of uncertainty in
the AB corrections for the SDSS photometric zeropoints.  
Figures \ref{fig:Tspec_ug} and \ref{fig:Tspec_gi} 
show the comparison of the DJE fits to the AB $u-g$ and $g-i$ colors.
Again, the agreement is good.

Based on the agreement of these three temperature estimates, we feel that there
is little doubt that the bulk of the stars in the sample are helium
atmosphere stars between 30,000~K and 45,000~K.  However, the small offsets
between the 3 methods speak to a mild level of systematic uncertainty
in the temperature scale, perhaps as much as 2000~K.  At higher signal-to-noise
ratio, there remain discrepencies between the spectroscopic and photometric
data.  Some of this may simply be due to be incorrect AB zeropoints, but we
doubt that this can be the full explanation.
If the residuals are the result of subtle differences in the line shapes,
this may be due to systematic errors in the models, but it might also be small 
problems in the SDSS fluxing corrections.

\subsection{Description of Individual Stars}
\label{sec:stars}

\subsubsection{Those with MMT Data}

\mystar{SDSS J001529.7$+$010521} 
This star has 11 epochs of photometry that prove that the star is very blue, 
with a color that suggests a temperature of 35,000~K.
We acquired spectroscopy from the MMT with the SPECPOL instrument.
The improved signal-to-noise ratio in the red allows us to exclude H$\alpha$,
and the spectrum is a good match to helium atmosphere at 35,500~K.
Hence, we regard this star as an extremely good case for a DB at 35--36,000K.

\mystar{SDSS J074538.1$+$312205} 
The SDSS spectrum of this star shows a hint of HeII 4686\AA\ line, and
so this star was reobserved at the MMT with Blue Channel.  
There is evidence for H$\beta$ in the MMT spectrum, and 
the HeII 4686\AA\ line is confirmed.
However, no atmosphere model fits the spectrum and so
we do not quote a temperature from the MMT spectrum.
We include the SDSS values, as they were part of the selection, but in light
of the MMT spectra, the fits to the SDSS spectra should not be trusted.
The photometry suggests $39,800\pm 2000$~K.  As mentioned in \S~\ref{sec:intro},
the presence of the HeII line requires a temperature
exceeding 38,000~K.  The relative strength of HeII 4686\AA\ compared 
to HeI 4713\AA\ is weak, suggesting that the star is
a cool DO, perhaps close to the photometric fit in temperature, 
which would make it the coolest DO star yet found.

Two explanations are suggested to explain this discrepant spectrum.
First, the star might have a weak magnetic field, enough to distort
the line profiles, making the cores more shallow, but not enough to
show Zeeman splitting.  Such was the case for LB~8827 = PG~0853+164
\citep{put97,wes01}.  This remarkable DBA with a probable temperature
moderately above 20,000~K shows variable circular polarization.  The
spectrum of SDSS J0745 might be monitored for time-dependent variations in the
He and H line strengths, modulated on a rotational period.  The
magnetic DBA Feige~7 shows evidence for spatial variation in the
He and H abundances modulated on its 2.2-hr rotational period
\citep{lie77,ach92}.
The second possibility is that the atmospheric composition might be
inhomogeneous, either because of stratification or because of spatial
variability around the surface.  This star would then be another 
example of the peculiar DAO stars mentioned in the introduction.

Thus, SDSS J074538.1$+$312205 might be observed with a spectropolarimeter
(on a {\it large} telescope) to search for weak circular polarization,
as found for LB~8827.  It might also be monitored for spectrum
variability, modulated on a rotational period.

\mystar{SDSS J090456.1$+$525030} 
SDSS spectrum was too noisy to exclude H$\alpha$, so we reobserved
this star at the MMT with the Blue Channel spectrograph.  
The photometry suggests $35,700\pm2000$ K.  The spectroscopic fits
to the MMT spectra are higher, around 40,000K.  However,
these fits are not good to the eye;
this is true even if one restricts the fit to blueward of 5400\AA.
The 40,000K model predicts a weak HeII 4686\AA\ line
that is absent from the MMT spectrum, and the line cores have the
wrong depth.  

The fits improve if one includes some hydrogen in the atmosphere.
The $10^{-2}$ H/He model grid produces a H$\alpha$ line that is
likely just permitted within the noise, and the agreement with the He line
depths improve.  There is a hint of a H$\beta$ line in the
spectrum that offers some support for this interpretation.
The fitted model temperature drops to 36,300~K.
The photometric fit drops to $34,200\pm2500$, compatible
with this.  In other words, a DBA fit around 36,000~K is 
a plausible outcome.  At present, we have left our classification
as DB pending a resolution of the discrepency.

\mystar{SDSS J095256.6$+$015407} 
The SDSS and MMT Blue Channel spectra are both of high quality and argue
against the presence of any Balmer lines.  The spectroscopic fits yield a 
temperature of 34,000K,
but this is about 2-$\sigma$ high for the photometry, which suggests
a temperature of 30,800K.

\mystar{SDSS J140159.1$+$022126} 
SDSS spectrum was too noisy to exclude H$\alpha$, so we reobserved
this star at the MMT with the Blue Channel spectrograph.  
The MMT spectrum suggests that H$\alpha$ and H$\beta$ are present.  
Fitting with a H/He ratio of 0.01 drops the spectroscopic fit
by 2400K, to 36,500K, relative to a pure-helium model fit.  
The photometric fit is compatible with this result.
Hence, although this star does have hydrogen in its atmosphere,
it is almost certainly above 30,000K.

\mystar{SDSS J141258.1$+$045602} 
The SDSS spectrum is very nice and does not show H$\alpha$.
This star was reobserved at the MMT with Blue Channel, and we 
again found no signs of Balmer lines.
The spectroscopic fit is about 31,500~K; the photometry is compatible
with this but somewhat redder, around 30,000~K.

\subsubsection{Stars with high $S/N$ SDSS data}

These four stars are bright enough that they have high quality SDSS data.

\mystar{SDSS J113609.5+484318} 
This is a bright DO+M binary system, such that the SDSS spectrum is of 
high signal-to-noise ratio.  The spectrum suggests a temperature of 
45,000~K to 46,400~K.  The $u-g$ and $g-r$ photometry is consistent with this; 
the $i$ and $z$ band photometry is anomalously bright because of the 
companion.  

\mystar{SDSS J154201.4$+$502532} 
The SDSS spectrum is very nice and shows no H$\alpha$.  The photometry
supports the spectroscopic fit at 32,000--33,000~K.

\mystar{SDSS J215514.4--075833} 
The autofit spectroscopic fits suggests a temperature of 32,000~K,
while the DK fit is about 3000~K cooler.  The photometry would seem
to support the higher temperature.
There is no sign of H$\alpha$ in the spectrum, but there is an unexplained
absorption feature at 4050\AA.

\mystar{SDSS J234709.3$+$001858} 
This star was observed twice by the SDSS at similar signal-to-noise ratio; 
the other spectrum is from plate 385, fiber 623, MJD 51783.  
Moreover, this star has 11 epochs
of photometry.  The spectra would suggest a temperature around 32,000--33,000~K, 
but the photometry is redder than that predicted by the model.  
The photometry would prefer 28,700~K. 
No Balmer lines are visible in the SDSS spectra.

\subsubsection{Other candidates from SDSS}

Most of the remaining objects should simply be considered
candidates.  The spectra are best explained as DB or DO white dwarfs
at high temperature and the photometry is generally blue.  
However, the spectra are generally too noisy to exclude weak Balmer lines
or to detect some of the peculiarities found in the above cases.
Of course, the temperature fits have more statistical uncertainty as well.

\mystar{SDSS J040854.6--043354} 
Some noisy hints of H$\alpha$ and H$\beta$, although these are suspiciously
narrow.  The spectroscopy is noisy but prefers 35,000--36,500~K even with a 1\%
hydrogen admixture.  The photometry would seem to be hotter, 40,000~K,
even with the hydrogen admixture, but this is disfavored by the lack
of a He~II~4686\AA\ detection.  The errors
are such that reconciliation closer to the spectroscopic values is the likely outcome.

\mystar{SDSS J081115.0+270621} 
This is a low temperature DO star, likely around 47,500~K.

\mystar{SDSS J081546.0+244603} 
This is a low temperature DO star, likely around 46,000~K.

\mystar{SDSS J084823.5+033216} 
The spectrum suggests 32,000--34,000~K and the photometry is 
consistent with this, if not a little bluer.

\mystar{SDSS J084916.1+013721} 
This star is likely at the low end of the temperature range
in this paper.  The spectrum is fit to 29,500~K, and the 
photometry is somewhat redder than this, albeit consistent.

\mystar{SDSS J090232.1+071929} 
The spectrum and photometry both favor 30,000~K.

\mystar{SDSS J092544.4+414803} 
The best fit to the spectrum is 39,000~K.  The photometry
is slightly redder than this, but consistent.

\mystar{SDSS J093041.8+011508} 
The spectrum is noisy but suggests 32,000-34,000~K.  However, 
the photometric fit is significantly cooler than this,
below 30,000~K with a best fit around 26,700~K.

\mystar{SDSS J093759.5+091653} 
The spectroscopy suggests 31,500~K while the photometry is 
about 2 $\sigma$ redder (27,600~K).

\mystar{SDSS J123750.4+085526} 
The spectroscopic fit is 31,000--34,000~K; the photometry is slightly
bluer but consistent.

\mystar{SDSS J134524.9--023714} 
The spectroscopic fits call for a 38,000-41,000~K temperature.
The photometry is somewhat redder but consistent.  The spectrum is
noisy enough that the HeII lines implied by the higher temperature
can't be excluded.  

\mystar{SDSS J141349.4+571716} 
The spectroscopy and photometry both support a 30,000-31,000~K temperature.
Although there is a small absorption feature near where
H$\alpha$ would be, the wavelength doesn't match exactly
and so we think this is probably noise.

\mystar{SDSS J143227.2+363215} 
The spectroscopic fits disagree somewhat, 29,000~K and 32,700~K,
but the higher temperature DK fit is clearly affected by continuum artifacts
that render it suspect to the eye.
The photometry argues for the lower value, with an upper limit of
about 31,000~K.

\mystar{SDSS J153852.3--012133} 
This star shows indications of H$\alpha$ and H$\beta$, 
so we classify it as DBA.  With 1\% hydrogen in the atmosphere,
the two spectroscopic fits fall on either side of the DBV degeneracy,
22,000~K and 30,000~K.  The dereddened photometry
is much bluer, suggesting 35,000~K and supporting the higher temperature.  
However, this line of 
sight has an unusually high extinction, $A_g=0.50$ mag, so 
it is possible that the photometry has been overcorrected.
While this star is probably around 30,000~K, the possibility that
it is on the lower temperature side of the degeneracy cannot 
be completely excluded with the SDSS data.

\mystar{SDSS J164703.4+245129} 
The spectrum is noisy, but the best fit is 32,000--33,000~K.  The photometry
is somewhat bluer than this model but consistent.

\mystar{SDSS J211149.5--053938} 
This star shows a hint of H$\alpha$,
so we classify it as DB A:.
The spectroscopic fit with a 1\% hydrogen atmosphere is 36,000~K with autofit;
the photometry supports this.  The DK fit finds 47,400~K, but this is not
consistent with the lack of HeII lines.  A higher quality red spectrum is 
required to confirm the H$\alpha$ signature.

\mystar{SDSS J212403.1+114230} 
The spectroscopic fits find 30,000--32,000~K, and the photometry is consistent
with this.

\mystar{SDSS J222833.8$+$141036} 
The spectrum shows indications of a faint M star companion redward of
7000\AA.  The photometry prefers a somewhat cooler temperature 
than the spectroscopy, 28,500~K versus 33,000~K, but some of
this skewing might be due to the $i$ band being made slightly
brighter by the companion.  The $u-g$ color, however, is consistent with the
spectroscopic temperature.

\section{Is there a DB gap?}

Having shown that the temperature range between 30,000~K and 45,000~K
is not devoid of helium atmosphere white dwarfs, we would like to
understand whether there is any deficiency at all of DB stars at
this temperature.  In other words, do all low-temperature DB stars go
through a high-temperature DB phase?  In principle, one would like to
construct a temperature function, namely the number density of stars
at each temperature, and compare that to the steady state distribution
given the cooling rates.  In practice, this is tricky because given
the flux limits of the SDSS, one is probing larger distances for the
hotter stars and these distances are large enough that we can see out
to several times the scale height of the Galactic disk.  In other words,
to construct the temperature function, one must first model the distribution 
of white dwarfs in the Galactic disk.

It is likely easier to use the DA stars as a sentinal for the Galactic
distribution by forming the ratio of the number of DA to DB stars in
the same volume of space and then studying the temperature dependence of
that ratio.  If DA and DB stars cool at the same rate, then in a given
volume of space if these two types of stars do not convert from one to
another (or convert into another class of white dwarf), then the ratio
should remain constant.  More generally, we simply require that the
two types of stars have cooling rates ($dT/dt$ as a function of $T$)
that differ by at most a temperature-independent multiple.
Comparison of hydrogen envelope and hydrogen-free cooling models
by G.~Fontaine in \citet{Lie05} indicates that no signficant
differences in the cooling rates occur for a mass of $0.6\msun$ 
between 45,000~K and 15,000~K, supporting our assumption.

We will use DA and DB white dwarfs from the SDSS DR4 white dwarf catalog
\citep{Eis05}, focusing on the region between 15,000 and 40,000~K.
For the hot DBs, we adopt the pure-helium fit to the SDSS spectra.
We optionally include DBA stars as DBs, without adjusting their temperatures for
the mixed atmosphere.  Otherwise, we exclude classes other than DB
or DA (e.g., we exclude DA+M binaries).  We adopt the DA temperatures
from \citet{Eis05}.  We do not use the errors on the temperatures, and 
we omit stars below 15,000~K because the helium lines are getting weak.

To make this a clean test, we want to probe the DA and DB stars in
the same volume of space (at each temperature).  It is acceptable for different
temperatures to probe different volumes so long as the comparison at each
temperature is fair.  A simple flux limit would nearly accomplish the
goal, but at a fixed temperature and surface gravity, DA and DB stars
do not have exactly the same optical magnitude because of the details
of the atmospheres and resulting bolometric corrections.  For example,
at 20,000~K a DA star is about 0.1 mag fainter in $g$ than a DB star,
whereas at 30,000~K, the DA star is about 0.25 mag brighter.  For our
analysis, we require at all temperatures that the DB stars have $g<19$.
We then find, at each temperature, the flux limit for the DA stars that
would detect both the DA and DB star out to the same distance.  Here,
we assume that both the DA and DB stars have the same surface gravity,
$\logg=8.0$.

Because DA and DB stars have a spread in radii and surface gravities,
a flux limit does not correspond to a fixed distance limit at a 
given temperature.  Larger stars (less massive white dwarfs) can be detected
to greater distance.  If DA
and DB stars had the same distribution of surface gravities, then
the fuzziness of the distance boundary would not be a problem.
However, it is known that DB stars have a tighter distribution of $\logg$ \citep{bea95}.
Populations of particularly high-mass and particularly low-mass DA stars
exist that don't seem to have a DB analog.  This is likely due to 
different formation mechanisms \citep[See discussion in][]{Lie05}.  To address this,
we only include DA stars with $7.5<\logg<8.3$.  We include all DB stars.
Note that although the autofit method does produce a surface gravity
estimate, we are not using this to compute a distance.

Finally, the completeness of the SDSS catalog must be modeled.  The 
important selection boundary is the $u-g<0$ cut of the Hot Standard
class.  DB stars are bluer than this at all temperatures above 14,000~K,
but DA stars become redder than $u-g=0$ around 22,000~K.  \citet{Eis05}
presents a simple completeness analysis for unblended blue stars, finding
that at $g<19$,
the completeness of the SDSS at $u-g>0$ is only 66\% of that at $u-g<0$.
We therefore upweight the redder stars by a factor of 1.5 and 
restrict ourselves to unblended stars where the completeness is highest.
The unblended cut itself produces an incompleteness that depends on 
magnitude and Galactic latitutde, but this is irrelevant for our comparison
because it affects DA and DB stars nearly equally.  We exclude blended
stars \citep{Eis05}.

\begin{figure}[p]
\centerline{\epsfxsize=5in\epsffile{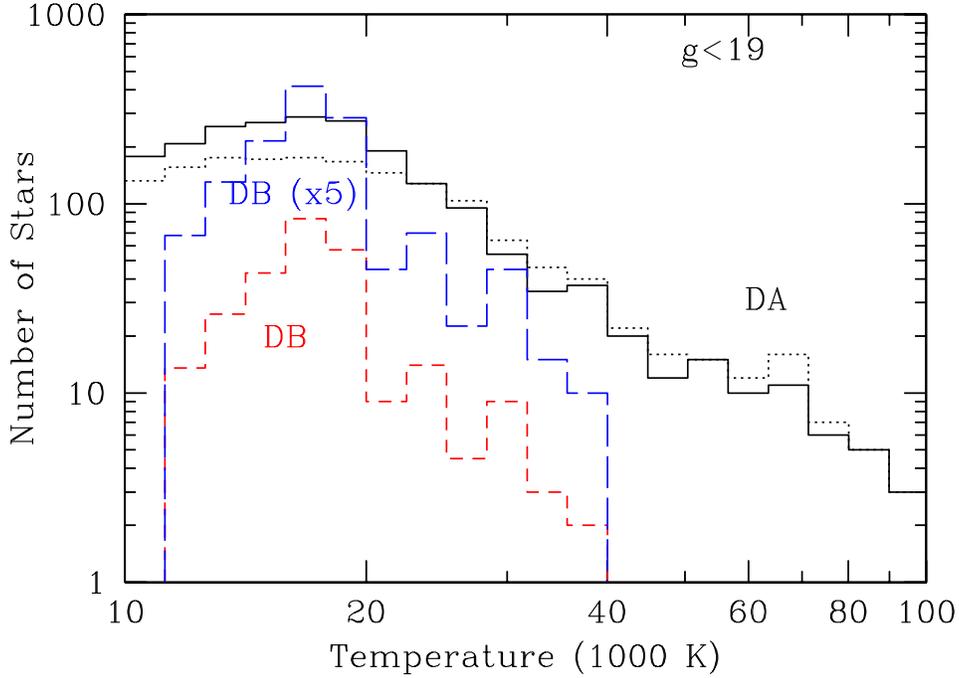}}
\caption{\label{fig:Thist}%
The temperature histogram of DA and DB stars from the DR4 catalog.
The short-dash red line shows the histogram of DB stars at $g<19$, 
including DBA stars.  The long-dashed blue line is the same histogram
multiplied by 5.  The dotted line is the histogram of DA stars to 
$g<19$.  The solid line is the histogram of DA stars, down to a 
temperature-dependent flux cut that matches the distance one can see
a DA star and a $g=19$ DB star of the same temperature and gravity.
The solid line has also been corrected for the extra incompleteness 
at $u-g>0$ and $g>19$.  One sees that the corrected DA histogram and
the $\times 5$ DB histogram match at 20,000~K, but that the DBs
have a factor of 2.5 shortfall at hotter temperatures.
The rolloff in the DA histogram below 15,000~K is likely due to 
the scale height of the Galactic disk.  The extra rolloff in the DB
distribution is because these stars become very weak-lined at
temperatures below about 14,000~K and hence are harder to identify.
}
\end{figure}

Figure \ref{fig:Thist} shows the histogram of temperatures for the DA and DB
stars, having weighted for color-dependent incompleteness and applying
the temperature-dependent flux cuts to the DAs.  Including the DBA stars,
the ratio of DA to DB
stars is 5 between 16,000 and 22,000~K (703 and 139 stars), but increases 
to 12.5 between 25,000 and 40,000~K (226 and 18 stars).  In other words, the 
DB gap appears to be deficient by a factor of 2.5.

Adjusting various aspects of our analysis does not appear to affect this
deficit.  Excluding the DBA stars or including the DAs of other $\logg$
alters the ratio of DA to DB stars, but the factor of roughly 2.5 between
the hot and cool ratios persists.  For example, excluding the DBA stars, the
factor is 2.2.  Similarly, if one assumes the DB stars
to be $\logg=8.5$ while leaving the DA stars at $\logg=8.0$, then the
flux limit for the DA sample becomes brighter to maintain the same volume,
but this simply increases the DB to DA ratio at all temperatures without
affecting the deficit in the gap.  In other words, the ratios 5 and 12.5
given above are less robust than their ratio.

Could this factor of 2.5 be an artifact of our temperature fits?  Such a result seems
difficult to arrange.  A relatively smooth systematic bias in either the
DA or DB fits would shift the histograms but wouldn't warp them enough to 
remove the deficit.  For the DA stars, the trends of the line profiles 
versus temperature over the range in question are strong and monotonic;
moreover, the results are supported by the locus of color versus temperature
\citep{Eis05}.  It is difficult to imagine how one could suppress the DA
histogram by a factor of 2.5 at 30,000~K.  For the DB stars, if the normalization
at 20,000~K is correct, then we should have found of order 70 hot DB stars
instead of the 28 in this paper.  We do not believe the catalog could be
this incomplete.  The final route is to try to soften the peak in the
DB histogram at 20,000~K and shift some of those stars towards lower
or higher temperatures.  But there are simply too many DB stars in the peak
(note that Fig.~\ref{fig:Thist} is on a logarithmic scale).  Even between
14,000 and 25,000~K, the ratio of DA to DB stars is 7, much less than 14,
and we don't plausibly see how to shift these rank-and-file DB stars to be
hotter than 25,000~K or less than 14,000~K.

If the change in DA to DB ratio from 30,000~K to 20,000~K is caused by
some DAs turning into DBs, and if the ratio of 5 at low temperatures
is correct, then this means that roughly 12\% of 30,000~K DA stars turn into
20,000~K DB stars.

If one considers yet hotter temperatures, then the DO stars seem to rebound
in number compared to the hot DAs.  The DA histogram includes 96 stars
at $T>40,000$~K down to $g=19$.  Meanwhile, the DR4 catalog includes 22 non-binary, non-blended
DO stars at $g<19$, most above 50,000~K.  There are 10 PG1159 stars as well,
and these are thought to mostly be white dwarfs, although some are
hot subdwarfs.
Hence, taking the simplest route, the ratio of DA to non-DA white dwarfs at 
temperatures above 40,000~K is about 3, less than even the value of 5 found
at 20,000~K and much less than the ratio at 30,000~K.
Because the temperature measurements in all three classes of these stars are
somewhat uncertain, we postpone a detailed treatment of the temperature
dependence of the ratio.  Moreover, we have not included any differential
bolometric correction for the survey depth for the DAs and DOs.  Still, 
there is little doubt that the DA fraction drops at the highest temperatures.

Although it is unlikely to matter, we should note that although
our calculations of DA to DB ratios is fair at each temperature, we
are probing different volumes of space as a function of temperature.
In other words, if the DB to DA ratio dropped as a function of distance
above the Galactic plane, we would see a trend with temperature that
could mimic an evolutionary transformation.  However, the fact that the
DB gap was discovered in much brighter and closer samples suggests that
this is not the cause.

Clearly, the above is only a first analysis, and more work is needed
to firm up this quantitative factor of 2.5 as the deficit in the DB gap and
to better measure the transition temperatures.  Such work could include
the surface gravities as a distance measure, but more importantly it
is necessary to more carefully place the DA and DB stars on the same
physical temperature and surface gravity scale.  Our autofit work can
only be considered approximate for this task.  
Figure \ref{fig:Thist} hints that the transition temperature may even be closer
to 20,000~K, but we have chosen to be conservative and not tune our bins
to fine details in the histogram. 
Further, to quantify the
gap more finely, one will probably need to consider what role the mixed types of
stars, such as DBA, or other classes of WD, such as the hot DQs, play
in the DB gap.  The numbers of these stars are sufficiently less than
the 55\% of the DB stars (i.e., the gap) that they can't plausibly fill
the gap, but their numbers might well be large enough to shift the quantification
of the deficiency.  Finally, subtle differences
in the cooling function between DA and DB could alter the amount
of transformation implied.  These kinds of detailed modeling and evolutionary
questions have been long-standing topics in the white dwarf literature.
Because of the complicated nature of these systematic uncertainties, we 
are hesitant to quote an error bar on our 2.5 value.  We expect that if
one pushed the numbers and the definitions, one could reach 2 or 3.
Hopefully this first quantification of the DB gap will encourage further work
on constraining these evolutionary paths.

\section{Conclusions}

In summary, we have presented 28 stars as candidate hot DB or cool DO
white dwarfs.  Some of the stars have compelling evidence for being
between 30,000 and 45,000~K, in the DB gap.  These are the first
helium-atmosphere white dwarfs found in this range of temperature.
Given the existence of strong proof in these cases, it is likely that
most of the candidates with noisier data are also at the temperatures
indicated by their spectroscopic and photometric fits.

Of the ten stars with particularly good observations,
six of them have strong evidence for being DB stars above 30,000 K:
SDSS J001529.7$+$010521,
SDSS J095256.6$+$015407,
SDSS J140159.1$+$022126 (despite some hydrogen),
SDSS J090456.1$+$525030 (but not a perfect fit), 
SDSS J154201.4$+$502532, and
SDSS J215514.43--075833.7.
Two other well-observed stars are only somewhat cooler, 28,000--30,000~K:
SDSS J141258.1$+$045602 and
SDSS J234709.3$+$001858. 
The final two are cool DO stars:
the DO+M star SDSS J113609.59+484318.9 at 45,000~K and
the possibly weakly magnetic SDSS J074538.17+312205.3.

The 18 other stars in this paper are noisier candidates from the same vein.
Two are cool DO stars, and one DBA (SDSS J153852.34--012133.7) is sufficiently
uncertain that could be below 25,000~K.  The other 15 are DB and DBA stars
with temperature estimates between 28,000~K and 39,000~K.
Some of these stars, notably SDSS J074538.1+312205, may be similar
to some of the previously known ``peculiar'' DAB/DBA/DAO stars discussed
above, but more observations will be needed to establish these links.

Helium atmospheres are not as well understood as hydrogen ones, and it is
certainly possible that the temperature scale inferred from our modeling
has systematic biases.  The trend among several of our examples with
better data for the spectroscopic fit to be $\sim2000$~K hotter than
the photometric fit may be an indication of such a bias.  On the other
hand, several of the stars have consistent results between the two
methods and some of the noisier examples have warmer photometric fits.
Whether these differences reflect real differences in the properties
of the atmospheres or simply unmodelled errors in the data remains
to be seen.  

Our results clearly show that helium atmosphere stars do exist at
temperatures above the DBV strip at $\sim\!25,000$ K and below the
HeII DO stars at 47,000~K and up.  The DB gap is not empty.  Indeed,
we find a continuous distribution of temperatures.  

However, in comparing the ratio of the numbers of DB and DA stars as
a function of temperature, we find that there remains a factor of 2.5
deficit in the density of DB stars at temperatures between 25,000
and 40,000~K as compared to a baseline established between 16,000
and 22,000~K.  This was a simple analysis, with several assumptions
about fine details that one would hope to improve, but our results
support the remarkable idea that about 10\% of 30,000~K DA stars become
DB stars by the time they have cooled to 20,000~K.  The mechanism for
this transformation remains a mystery.

\bigskip

DJE was supported by grant AST-0098577 and AST-0407200 from the National Science
Foundation and by an Alfred P. Sloan Research Fellowship.
JL acknowledges support from NSF grant AST-0307321 for work on 
white dwarfs found in the SDSS. 

Funding for the creation and distribution of the SDSS Archive has
been provided by the Alfred P. Sloan Foundation, the Participating
Institutions, the National Aeronautics and Space Administration,
the National Science Foundation, the U.S. Department of Energy, the
Japanese Monbukagakusho, and the Max Planck Society. The SDSS Web site
is http://www.sdss.org/.

The SDSS is managed by the Astrophysical Research Consortium (ARC)
for the Participating Institutions. The Participating Institutions are
The University of Chicago, Fermilab, the Institute for Advanced Study,
the Japan Participation Group, The Johns Hopkins University, the Korean Scientist Group,
Los Alamos
National Laboratory, the Max-Planck-Institute for Astronomy (MPIA), the
Max-Planck-Institute for Astrophysics (MPA), New Mexico State University,
University of Pittsburgh, University of Portsmouth, Princeton University, the United States Naval
Observatory, and the University of Washington.

\end{document}

%% file: db_table1.dat
SDSS J001529.74+010521.3 & DB & 389 & 530 & 51795 & 2662 & 40 & 6 & 308 & 77 & Y & 3.873926 & 1.089256 & 2.08 & 184.6 \\
SDSS J040854.60--043354.6 & DB A: & 465 & 518 & 51910 & 1729 & 40 & 6 & 496 & 29 & Y & 62.227510 & --4.565178 & 1.79 & 183.4 \\
SDSS J074538.17+312205.3 & DO & 755 & 51 & 52235 & 2137 & 40 & 2 & 91 & 28 & Y & 116.409046 & 31.368161 & 3.57 & 261.7 \\
SDSS J081115.08+270621.7 & DO & 1206 & 235 & 52670 & 3462 & 40 & 4 & 86 & 192 & Y & 122.812863 & 27.106047 & 1.35 & 185.2 \\
SDSS J081546.08+244603.2 & DO & 1585 & 388 & 52962 & 3644 & 41 & 5 & 96 & 30 & N & 123.942021 & 24.767581 & 0.30 & 338.7 \\
SDSS J084823.52+033216.7 & DB & 564 & 565 & 52224 & 2125 & 40 & 5 & 89 & 145 & N & 132.098019 & 3.537982 & 0.55 & 160.6 \\
SDSS J084916.18+013721.2 & DB & 467 & 594 & 51901 & 1462 & 40 & 4 & 46 & 188 & N & 132.317420 & 1.622575 & 1.74 & 172.0 \\
SDSS J090232.17+071929.9 & DB & 1191 & 368 & 52674 & 3478 & 40 & 5 & 129 & 133 & Y & 135.634078 & 7.324996 & 1.82 & 286.5 \\
SDSS J090456.11+525029.8 & DB & 552 & 547 & 51992 & 2074 & 41 & 6 & 125 & 46 & Y & 136.233828 & 52.841617 & 1.69 & 246.1 \\
SDSS J092544.40+414803.1 & DB & 939 & 362 & 52636 & 2887 & 40 & 6 & 225 & 45 & N & 141.435022 & 41.800881 & 2.20 & 269.3 \\
SDSS J093041.80+011508.4 & DB & 475 & 101 & 51965 & 1907 & 40 & 2 & 134 & 47 & Y & 142.674168 & 1.252336 & 0.46 & 180.1 \\
SDSS J093759.52+091653.2 & DB & 1304 & 131 & 52993 & 3538 & 40 & 3 & 178 & 30 & Y & 144.498002 & 9.281449 & 1.78 & 218.1 \\
SDSS J095256.68+015407.6 & DB & 481 & 513 & 51908 & 1907 & 40 & 3 & 171 & 55 & Y & 148.236183 & 1.902138 & 1.25 & 150.8 \\
SDSS J113609.59+484318.9 & DO+M & 966 & 69 & 52642 & 3059 & 42 & 2 & 105 & 30 & Y & 174.039961 & 48.721934 & 1.06 & 169.4 \\
SDSS J123750.46+085526.0 & DB & 1233 & 83 & 52734 & 3525 & 40 & 1 & 229 & 49 & Y & 189.460269 & 8.923910 & 0.14 & 157.2 \\
SDSS J134524.92--023714.2 & DB & 913 & 185 & 52433 & 2333 & 40 & 3 & 63 & 134 & N & 206.353845 & --2.620620 & 0.77 & 111.9 \\
SDSS J140159.09+022126.7 & DB & 532 & 358 & 51993 & 1462 & 40 & 4 & 568 & 58 & Y & 210.496233 & 2.357431 & 0.76 & 345.3 \\
SDSS J141258.17+045602.2 & DB & 583 & 432 & 52055 & 1478 & 40 & 5 & 35 & 54 & Y & 213.242391 & 4.933957 & 2.69 & 168.0 \\
SDSS J141349.46+571716.4 & DB & 1160 & 414 & 52674 & 3183 & 40 & 5 & 110 & 87 & Y & 213.456085 & 57.287893 & 0.46 & 333.4 \\
SDSS J143227.25+363215.2 & DB & 1382 & 314 & 53115 & 3893 & 41 & 1 & 349 & 60 & N & 218.113551 & 36.537568 & 2.14 & 353.0 \\
SDSS J153852.34--012133.7 & DBA & 926 & 428 & 52413 & 2334 & 40 & 4 & 259 & 214 & Y & 234.718125 & --1.359367 & 2.82 & 152.6 \\
SDSS J154201.48+502532.0 & DB & 796 & 180 & 52401 & 2299 & 41 & 3 & 252 & 130 & N & 235.506191 & 50.425559 & 2.68 & 296.9 \\
SDSS J164703.44+245129.0 & DB & 1414 & 445 & 53135 & 3958 & 40 & 3 & 163 & 86 & N & 251.764343 & 24.858082 & 1.81 & 176.0 \\
SDSS J211149.59--053938.3 & DB A: & 638 & 533 & 52081 & 1659 & 40 & 6 & 85 & 67 & Y & 317.956629 & --5.660647 & 0.93 & 156.9 \\
SDSS J212403.12+114230.1 & DB & 730 & 380 & 52466 & 1739 & 40 & 5 & 48 & 348 & Y & 321.013001 & 11.708370 & 2.32 & 112.0 \\
SDSS J215514.43--075833.7 & DB & 716 & 344 & 52203 & 2576 & 40 & 4 & 49 & 108 & Y & 328.810149 & --7.976054 & 2.46 & 207.4 \\
SDSS J222833.82+141036.9 & DB+M: & 737 & 602 & 52518 & 2566 & 40 & 6 & 161 & 68 & Y & 337.140934 & 14.176924 & 2.09 & 159.3 \\
SDSS J234709.29+001858.0 & DB & 385 & 624 & 51877 & 2662 & 40 & 4 & 261 & 80 & N & 356.788735 & 0.316125 & 1.11 & 43.1 \\

%% file: db_table2.dat
SDSS J001529.74+010521.3 & 18.711(26)  & 18.938(37)  & 19.419(22)  & 19.730(28)  & 20.09(12)  & 0.094 \\
SDSS J040854.60--043354.6 & 19.042(26)  & 19.307(14)  & 19.688(26)  & 19.964(40)  & 20.14(15)  & 0.364 \\
SDSS J074538.17+312205.3 & 18.592(19)  & 18.912(15)  & 19.348(22)* & 19.637(22)  & 19.89(08)  & 0.195 \\
SDSS J081115.08+270621.7 & 17.750(16)  & 18.118(14)  & 18.666(21)  & 18.999(16)  & 19.38(07)  & 0.111 \\
SDSS J081546.08+244603.2 & 18.170(21)* & 18.595(20)  & 19.045(21)  & 19.398(30)  & 19.65(08)  & 0.183 \\
SDSS J084823.52+033216.7 & 19.207(27)  & 19.446(17)  & 19.864(22)  & 20.242(36)  & 20.47(14)  & 0.129 \\
SDSS J084916.18+013721.2 & 19.296(30)  & 19.452(21)  & 19.802(32)  & 20.048(37)  & 20.54(22)  & 0.141 \\
SDSS J090232.17+071929.9 & 18.543(20)  & 18.711(18)  & 19.050(23)  & 19.376(29)  & 19.48(07)  & 0.260 \\
SDSS J090456.11+525029.8 & 18.665(22)  & 18.955(23)  & 19.411(17)  & 19.740(25)  & 19.93(09)  & 0.055 \\
SDSS J092544.40+414803.1 & 19.128(32)  & 19.303(38)  & 19.852(41)  & 20.310(57)  & 20.29(14)  & 0.059 \\
SDSS J093041.80+011508.4 & 19.168(28)  & 19.215(27)  & 19.614(28)  & 19.808(35)  & 20.00(12)  & 0.272 \\
SDSS J093759.52+091653.2 & 18.333(18)  & 18.520(20)  & 18.849(23)  & 19.184(33)  & 19.31(07)  & 0.154 \\
SDSS J095256.68+015407.6 & 17.292(24)  & 17.496(13)  & 17.849(19)  & 18.156(15)  & 18.45(03)  & 0.189 \\
SDSS J113609.59+484318.9 & 16.430(28)  & 16.836(33)  & 17.332(21)  & 17.513(20)  & 17.39(02)  & 0.073 \\
SDSS J123750.46+085526.0 & 18.618(34)  & 18.840(22)  & 19.314(28)  & 19.619(48)  & 19.96(15)  & 0.090 \\
SDSS J134524.92--023714.2 & 19.213(31)  & 19.467(20)  & 19.916(26)  & 20.202(37)  & 20.32(14)  & 0.173 \\
SDSS J140159.09+022126.7 & 18.736(28)  & 19.026(26)  & 19.470(25)  & 19.828(34)  & 19.99(13)  & 0.120 \\
SDSS J141258.17+045602.2 & 17.191(13)  & 17.346(23)  & 17.750(15)  & 18.086(20)  & 18.39(03)  & 0.079 \\
SDSS J141349.46+571716.4 & 18.226(55)  & 18.399(24)  & 18.808(19)  & 19.106(22)  & 19.51(09)  & 0.051 \\
SDSS J143227.25+363215.2 & 18.510(17)  & 18.653(24)  & 19.116(14)  & 19.378(22)  & 19.68(07)  & 0.038 \\
SDSS J153852.34--012133.7 & 19.307(33)  & 19.483(20)  & 19.791(21)  & 20.040(31)  & 20.39(18)  & 0.502 \\
SDSS J154201.48+502532.0 & 16.498(16)* & 16.777(23)  & 17.137(15)  & 17.502(15)  & 17.81(02)  & 0.066 \\
SDSS J164703.44+245129.0 & 19.219(29)  & 19.465(21)  & 19.854(23)  & 20.175(35)  & 20.47(13)  & 0.192 \\
SDSS J211149.59--053938.3 & 18.644(22)  & 18.839(12)  & 19.228(18)  & 19.492(24)  & 19.61(08)  & 0.415 \\
SDSS J212403.12+114230.1 & 18.881(23)  & 18.995(12)  & 19.329(17)  & 19.636(30)  & 19.93(11)  & 0.298 \\
SDSS J215514.43--075833.7 & 18.072(21)  & 18.270(27)  & 18.715(22)  & 18.992(28)  & 19.23(07)  & 0.115 \\
SDSS J222833.82+141036.9 & 18.611(25)  & 18.769(17)  & 19.084(18)  & 19.380(23)  & 19.66(08)  & 0.248 \\
SDSS J234709.29+001858.0 & 18.781(38)* & 18.997(27)  & 19.388(17)  & 19.656(24)  & 20.07(13)  & 0.107 \\

%% file: db_table3.dat
SDSS J001529.74+010521.3$^b$ & --0.336 & --0.820 & 35.4 & 33.6--37.3 & 0.36 & 4 \\ 
SDSS J040854.60--043354.6 & --0.437 & --0.836 & 40.1$^c$ & 34.2--45.7 & 1.29 & 4 \\ 
SDSS J074538.17+312205.3 & --0.431 & --0.828 & 39.8 & 35.4--44.1 & 2.00 & 3 \\ 
SDSS J081115.08+270621.7 & --0.448 & --0.946 & 47.4 & 44.7--~~~~~ & 4.05 & 4 \\ 
SDSS J081546.08+244603.2 & --0.531 & --0.900 & 43.6 & 35.1--~~~~~ & 0.47 & 3 \\ 
SDSS J084823.52+033216.7 & --0.326 & --0.869 & 35.5 & 31.3--40.5 & 2.01 & 4 \\ 
SDSS J084916.18+013721.2 & --0.247 & --0.674 & 25.8 & 19.9--29.8 & 2.12 & 4 \\ 
SDSS J090232.17+071929.9 & --0.302 & --0.797 & 30.7 & 27.4--34.3 & 4.42 & 4 \\ 
SDSS J090456.11+525029.8 & --0.350 & --0.825 & 35.7 & 31.9--40.0 & 1.14 & 4 \\ 
SDSS J092544.40+414803.1 & --0.236 & --1.049 & 37.1 & 31.7--43.6 & 11.84 & 4 \\ 
SDSS J093041.80+011508.4 & --0.185 & --0.730 & 26.7 & 21.9--30.5 & 3.43 & 4 \\ 
SDSS J093759.52+091653.2 & --0.283 & --0.748 & 27.6 & 24.3--31.1 & 4.82 & 4 \\ 
SDSS J095256.68+015407.6 & --0.312 & --0.760 & 30.8 & 28.0--33.8 & 2.53 & 4 \\ 
SDSS J113609.59+484318.9 & --0.472 & --0.725 & 46.1 & 40.7--~~~~~ & 0.06 & 2$^d$ \\ 
SDSS J123750.46+085526.0 & --0.295 & --0.835 & 34.8 & 30.1--40.7 & 1.10 & 4 \\ 
SDSS J134524.92--023714.2 & --0.357 & --0.828 & 36.9 & 32.2--42.4 & 1.82 & 4 \\ 
SDSS J140159.09+022126.7 & --0.373 & --0.871 & 37.6$^c$ & 32.0--43.7 & 1.30 & 4 \\ 
SDSS J141258.17+045602.2 & --0.224 & --0.791 & 29.7 & 27.0--32.4 & 3.74 & 4 \\ 
SDSS J141349.46+571716.4 & --0.231 & --0.745 & 28.4 & 24.3--32.9 & 1.74 & 4 \\ 
SDSS J143227.25+363215.2 & --0.197 & --0.757 & 28.1 & 25.3--31.1 & 4.13 & 4 \\ 
SDSS J153852.34--012133.7 & --0.398 & --0.798 & 34.9$^c$ & 29.7--41.1 & 1.47 & 4 \\ 
SDSS J154201.48+502532.0 & --0.343 & --0.770 & 31.5 & 27.0--37.7 & 7.18 & 3 \\ 
SDSS J164703.44+245129.0 & --0.355 & --0.811 & 35.0 & 30.7--40.0 & 0.63 & 4 \\ 
SDSS J211149.59--053938.3 & --0.385 & --0.855 & 38.5$^c$ & 33.8--43.4 & 2.39 & 4 \\ 
SDSS J212403.12+114230.1 & --0.262 & --0.790 & 29.9 & 26.7--33.4 & 1.68 & 4 \\ 
SDSS J215514.43--075833.7 & --0.280 & --0.789 & 31.8 & 28.3--35.5 & 1.20 & 4 \\ 
SDSS J222833.82+141036.9 & --0.288 & --0.738 & 28.5 & 25.3--31.9 & 2.64 & 4 \\ 
SDSS J234709.29+001858.0$^b$ & --0.229 & --0.767 & 28.7 & 27.2--30.2 & 3.04 & 4 \\ 

%% file: db_table4.dat
SDSS J001529.74+010521.3  & DB    & 11.6 & 36.7(08) & 7.97(12) & 1.07 & 40.0(08) & 8.31(10) & 1.28 \\
\hspace*{\fill}MMT & \nodata & \nodata & 35.5 & 7.68 & \nodata & 35.4 & 8.06 & \nodata \\
SDSS J040854.60--043354.6 & DB A: & 10.6 & 34.9(17)$^\dag$ & 7.61(16)$^\dag$ & 1.22 & 36.5(14)$^\dag$ & 7.82(12)$^\dag$ & 1.67 \\
SDSS J074538.17+312205.3  & DO    &  8.5 & 44.5(11) & 8.61(26) & 0.93 & 47.2(05) & 7.73(15) & 1.34 \\
\hspace*{\fill}MMT & \nodata & \nodata & \multicolumn{2}{c}{Poor Fit} & \nodata & \multicolumn{2}{c}{Poor Fit} & \nodata \\[3pt]
SDSS J081115.08+270621.7  & DO    & 16.3 & 47.5(01) & 8.01(04) & 0.96 & 47.4(03) & 7.78(08) & 1.29 \\
SDSS J081546.08+244603.2  & DO    & 15.7 & 45.5(04) & 7.72(09) & 0.95 & 46.5(01) & 7.84(08) & 1.21 \\
SDSS J084823.52+033216.7  & DB    & 11.1 & 32.1(07) & 7.70(10) & 1.03 & 34.3(07) & 7.55(09) & 1.31 \\
SDSS J084916.18+013721.2  & DB    &  8.4 & 29.4(10) & 7.72(13) & 1.09 & 29.5(07) & 7.36(12) & 1.38 \\
SDSS J090232.17+071929.9  & DB    & 12.9 & 30.2(06) & 7.71(08) & 0.99 & 29.9(05) & 7.50(08) & 1.20 \\[3pt]
SDSS J090456.11+525029.8  & DB    & 15.5 & 37.5(06) & 7.99(09) & 1.09 & 38.2(06) & 7.90(08) & 1.07 \\
\hspace*{\fill}MMT & \nodata & \nodata & 40.1 & 7.91 & \nodata & 39.4 & 7.85 & \nodata \\
SDSS J092544.40+414803.1  & DB    &  9.5 & 38.7(11) & 7.59(14) & 1.15 & 39.8(11) & 7.30(14) & 1.34 \\
SDSS J093041.80+011508.4  & DB    &  8.1 & 31.9(09) & 7.72(15) & 0.96 & 34.0(11) & 7.50(15) & 1.45 \\
SDSS J093759.52+091653.2  & DB    & 17.4 & 31.5(04) & 8.10(07) & 0.97 & 31.3(04) & 7.92(05) & 1.00 \\[3pt]
SDSS J095256.68+015407.6  & DB    & 30.6 & 33.8(02) & 8.16(03) & 0.82 & 34.8(02) & 8.07(03) & 1.19 \\
\hspace*{\fill}MMT & \nodata & \nodata & 34.2 & 8.30 & \nodata & 33.9 & 8.10 & \nodata \\
SDSS J113609.59+484318.9  & DO+M  & 28.9 & 45.0(01) & 7.62(02) & 0.99 & 46.4(00) & 7.81(03) & 1.22 \\
SDSS J123750.46+085526.0  & DB    &  9.3 & 31.3(08) & 7.82(11) & 0.94 & 34.0(06) & 7.87(10) & 1.44 \\
SDSS J134524.92--023714.2 & DB    &  8.0 & 38.2(14) & 7.90(17) & 1.07 & 41.4(14) & 7.46(17) & 1.50 \\[3pt]
SDSS J140159.09+022126.7  & DB    & 10.4 & 34.8(14)$^\dag$ & 7.98(15)$^\dag$ & 1.16 & 37.9(15)$^\dag$ & 8.01(13)$^\dag$ & 1.28 \\
\hspace*{\fill}MMT & \nodata & \nodata & 36.5$^\dag$ & 7.67$^\dag$ & \nodata & 36.5$^\dag$ & 7.51$^\dag$ & \nodata \\
SDSS J141258.17+045602.2  & DB    & 28.2 & 31.0(02) & 7.98(03) & 0.86 & 31.9(02) & 8.00(03) & 1.25 \\
\hspace*{\fill}MMT & \nodata & \nodata & 31.5 & 7.96 & \nodata & 31.2 & 7.87 & \nodata \\
SDSS J141349.46+571716.4  & DB    & 15.5 & 30.0(05) & 7.92(07) & 0.96 & 31.3(04) & 7.96(06) & 1.24 \\[3pt]
SDSS J143227.25+363215.2  & DB    & 12.0 & 29.1(06) & 7.57(09) & 1.19 & 32.7(05) & 7.82(09) & 1.33 \\
SDSS J153852.34--012133.7 & DBA   &  6.8 & 21.9(03)$^\dag$ & 8.49(15)$^\dag$ & 1.07 & 30.1(16)$^\dag$ & 8.26(16)$^\dag$ & 1.40 \\
SDSS J154201.48+502532.0  & DB    & 36.3 & 32.4(02) & 7.83(03) & 0.80 & 33.2(01) & 7.84(02) & 1.36 \\
SDSS J164703.44+245129.0  & DB    &  7.8 & 31.7(10) & 8.03(14) & 1.04 & 33.2(09) & 7.79(14) & 1.07 \\
SDSS J211149.59--053938.3 & DB A: &  8.1 & 36.1(22)$^\dag$ & 7.62(19)$^\dag$ & 1.00 & 47.4(07)$^\dag$ & 7.73(18)$^\dag$ & 1.34 \\[3pt]
SDSS J212403.12+114230.1  & DB    & 12.8 & 30.3(06) & 7.92(08) & 1.07 & 32.4(05) & 8.08(08) & 1.13 \\
SDSS J215514.43--075833.7 & DB    & 19.5 & 32.1(04) & 8.25(05) & 1.32 & 28.8(04) & 7.79(05) & 1.55 \\
SDSS J222833.82+141036.9  & DB+M: & 12.4 & 33.2(07) & 8.02(10) & 1.04 & 33.6(05) & 7.91(08) & 1.22 \\
SDSS J234709.29+001858.0  & DB    & 11.6 & 32.1(07) & 7.92(10) & 1.25 & 32.9(07) & 7.80(11) & 1.01 \\